\documentclass[preprints,article,accept,moreauthors,pdftex]{Definitions/mdpi}

\firstpage{1} 
\pubvolume{xx}
\issuenum{1}
\articlenumber{5}
\pubyear{2020}
\copyrightyear{2020}
\history{Received: date; Accepted: date; Published: date}

\pdfoutput=1

\usepackage{amssymb,amsmath}
\usepackage{mathtools}
\usepackage{dsfont}
\usepackage{subcaption}
\usepackage{algorithm}
\usepackage{todonotes}
\usepackage{algorithmic}
\usepackage{nicefrac}
\usepackage{units}
\newcommand{\R}{\mathds{R}}
\newcommand{\N}{\mathds{N}}

\newcommand{\Var}[1]{\text{Var}\left[ #1 \right]}

\newcommand{\E}[1]{\text{E}\left[ #1 \right] }
\renewcommand{\P}[1]{\text{Pr}\left[ #1 \right]}

\newcommand{\DesignN}{\underline{\xi}_N}

\newcommand{\Design}{\xi}

\newcommand{\Fim}{M}

\newcommand{\DESIGN}[1]{\Xi\left( #1 \right)}

\newcommand{\Criteria}{\Phi}

\Title{An Adaptive Algorithm based on High-Dimensional Function Approximation to obtain Optimal Designs.}

\Author{Philipp Seufert $^{1}$*, Jan Schwientek $^{1}$ and Michael Bortz $^{1}$}

\AuthorNames{Philipp Seufert, Jan Schwientek and Michael Bortz}

\address[1]{%
$^{1}$ \quad Fraunhofer Center for Machine Learning and ITWM, Fraunhofer-Platz 1, 67663 Kaiserslautern, Germany; jan.schwientek@itwm.fraunhofer.de (J.S); michael.bortz@itwm.fraunhofer.de (M.B.)}

\corres{Correspondence: philipp.seufert@itwm.fraunhofer.de; Tel.: +49-631-31600-4062}

\abstract{Algorithms which compute locally optimal continuous designs often rely on a finite design space or on repeatedly solving a complex non-linear program. Both methods require extensive evaluations of the Jacobian $Df$ of the underlying model. These evaluations present a heavy computational burden. Based on the Kiefer-Wolfowitz Equivalence Theorem we present a novel design of experiments algorithm which computes optimal designs in a continuous design space. For this iterative algorithm we combine an adaptive Bayes-like sampling scheme with Gaussian process regression to approximate the directional derivative of the design criterion. The approximation allows us to adaptively select new design points on which to evaluate the model. The adaptive selection of the algorithm requires significantly less evaluations of $Df$ and reduces the runtime of the computations. We show the viability of the new algorithm on two examples from chemical engineering.}


\begin{document}


\section{Introduction}\label{sec:Introduction}

In chemical engineering, the use of models is indispensable to describe, design and optimize processes - both on a lab and on production scales, both with academic and with industrial backgrounds. However, each model prediction is only as good as the model - which means that the reliability of models to describe and predict the outcome of real-world processes is crucial.

A precise model $f \colon X \to Y$ gives a good understanding of the underlying phenomenon and is the basis for reliable simulation and optimization results. These models  often depend on a variety of parameters $\theta$ which need to be estimated from measured data. Therefore experiments are performed and measurements taken in order to obtain a good estimate. Ideally, these experiments should be as informative as possible, such that the estimate of the model parameters is most accurate within the measurement errors.

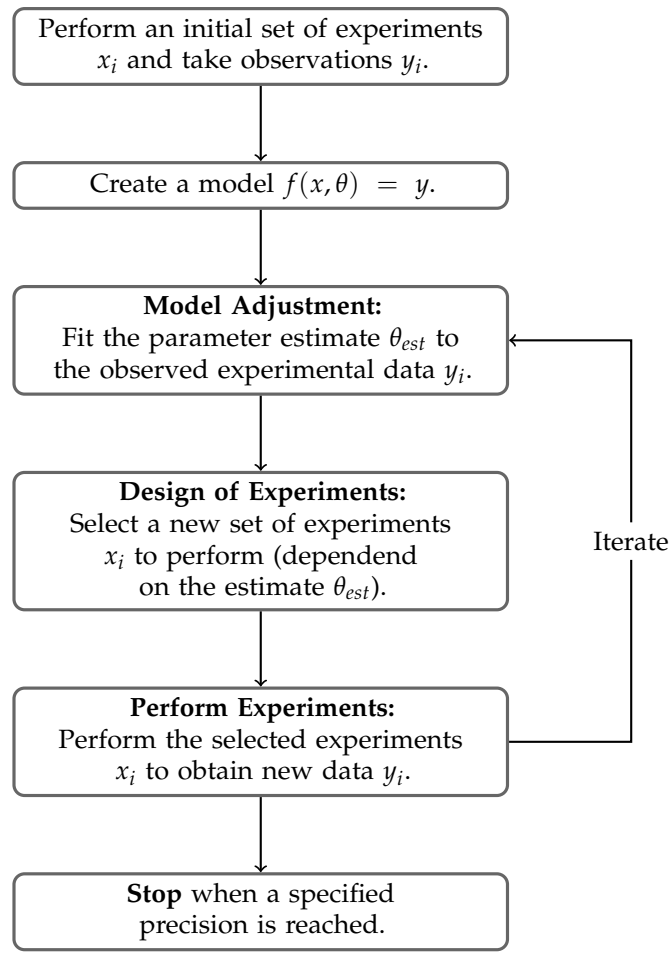
\begin{figure}[H]
\centering
\begin{tikzpicture}[squarednode/.style={rectangle, rounded corners, text centered, draw=black!60, text width=18em, minimum height=2em, very thick, minimum size=5mm},]
\node[squarednode] (InitialExperiments) {Perform an initial set of experiments $x_i$ and take observations $y_i$.};
\node[squarednode] (InitialModel) [below=of InitialExperiments] {Create a model $f(x,\theta) = y$.};
\node[squarednode] (ModelAdjustment) [below=of InitialModel] {\textbf{Model Adjustment:} \\ Fit the parameter estimate $\theta_{est}$ to the observed experimental data $y_i$.};
\node[squarednode] (DoE) [below=of ModelAdjustment] {\textbf{Design of Experiments:} \\ Select a new set of experiments $x_i$ to perform (dependend on the estimate $\theta_{est}$).};
\node[squarednode] (Experiments) [below=of DoE] {\textbf{Perform Experiments:} \\ Perform the selected experiments $x_i$ to obtain new data $y_i$.};
\node[squarednode] (Stop) [below=of Experiments] {\textbf{Stop} when a specified precision is reached.};
\node (ArrowText) [right=of DoE] {Iterate};
\draw[thick,->] (InitialExperiments.south) -- (InitialModel.north);
\draw[thick,->] (InitialModel.south) -- (ModelAdjustment.north);
\draw[thick,->] (ModelAdjustment.south) -- (DoE.north);
\draw[thick,->] (DoE.south) -- (Experiments.north);
\draw[thick,->] (Experiments.south) -- (Stop.north);
\draw [thick,-] (Experiments.east)  -|   (ArrowText.south);
\draw [thick,->] (ArrowText.north)  |-   (ModelAdjustment.east);
\end{tikzpicture}
\caption{Iterative Model Identification Workflow.}
\label{fig:Workflow}
\end{figure}

One approach often considered is an iterative workflow, which is presented in Figure \ref{fig:Workflow}. We begin by performing an initial set of experiments and measure corresponding results. With this data we construct a model $f(x,\theta)$ which describes the experiments. We also compute an estimate $\theta_{est}$ of the unknown parameters $\theta$ (\emph{model adjustment}). The next step is to find a new set of experiments to perform. Finding appropriate experiments is subject of \emph{model-based design of experiments} (M-bDoE or DOE). Then we perform the new experiments and collect new data. We iterate the process of adjusting the model parameters, selecting experiments and performing experiments until we are satisfied with the parameter estimate $\theta_{est}$.

In model based design of experiments we want to use the current information of our model in order to find optimal experiments. Here optimal means finding experiments which give the most information on the unknown parameters $\theta$. Equivalently, we want to minimize the error of our estimate as much as possible \cite{FederovBook,Atkinson,Aspiron2020,YLIRUKA2019637}.


The designs we compute will depend on the information we currently have on our model. In particular we have a dependency on our current estimate $\theta_{est}$ of the unknown parameters $\theta$. Such designs are called \emph{locally optimal}. In contrast, one can also compute \emph{robust} optimal designs which take uncertainty of the current estimate $\theta_{est}$ into account \cite{AspreyRobust, BarzUncertainty, KoerkelRobust}. Such robust designs are often used in the initial iteration, when no estimate of the parameters is given or the estimate is assumed to contain large errors. In later iterations of the workflow the estimate will be more precise and the locally optimal design is more reliable. Designs which are robust in the parameters however will not be subject of this paper and we instead consider locally optimal designs. 


The computation of locally optimal designs proves to be very challenging, as we have to consider integer variables - how many experiments to perform - as well as continuous variables - which experiments to perform. One can therefore solve the optimal design problem for continuous designs, which do not depend on the number of experiments \cite{FederovBook,Atkinson,boyd_vandenberghe_2004,Vanaret2020,Schwientek2020} and instead assign each experiment $x_i$ a weight $w_i$. The design points with strictly positive weights $w_i>0$ then indicate which experiments to perform, whereas the size of the weights indicates how often to perform each experiment. 


Classical algorithms to compute optimal continuous designs include the Vector Direction Method by Wynn and Federov \cite{WynnAlgorithm, FederovBook}, a Vector Exchange Method by Boehning \cite{BoehningVEM} and a Mulitplicative Algorithm by Tritterington \cite{TritteringtonMUL}. These methods are based on the heralded Kiefer-Wolfowitz Equivalence Theorem \cite{FederovBook,KieferEquivalenceTheorem} and compute locally optimal continuous designs. 

In recent years a variety of new algorithms were developed in order to compute optimal continuous designs. Prominent examples include the Cocktail Algorithm by Yu \cite{YuCocktail,YuConvergence}, the YBT Algorithm by Yang, Biedermann and Tang \cite{YBTAlgorithmus}, an adaptive grid algorithm by Duarte \cite{DuarteSDP} or the Random Exchange Algorithm \cite{HarmanRandExchangeAlg}.  These algorithms also rely on the Kiefer-Wolfowitz Equivalence Theorem. Each algorithm requires the repeated solution of the optimization problem

\begin{equation*}
    x^* = \arg \min_{x \in X} \phi(\Design,x).
\end{equation*}
Here $\phi(\Design,x)$ denotes the directional derivative \cite{YBTAlgorithmus} of the design criterion and is in general a non-linear function.

Finding global solutions to this problem is very challenging. One approach is to replace the design space $X$ with a finite grid $X_{grid}$ and solve the optimization via brute-force \cite{YBTAlgorithmus,HarmanRandExchangeAlg,YuCocktail}. This works very well for `small' input spaces and models. However, in applications we often find high-dimensional models, which are additionally time expensive to evaluate. In particular dynamical models fall under this category. Here, control or input functions need to be parametrized in order to be representable in the DoE setting. Depending on their detail, these parametrizations can result in a large amount of input variables.

For such high-dimensional models a grid-based approach is not viable. To compute the values $\phi(\Design,x)$ we need the Fisher Information Matrix $\mu(x)$ at every design point $x$ of the grid $X_{grid}$. The Fisher Information Matrix depends on the Jacobian $Df$ of the model $f$ and is thus time-expensive to compute. The time for evaluating these matrices on a fine high-dimensional grid scales exponentially with the dimension and therefore eventually will be computationally too expensive. Thus we need a different approach for these models. 

One should note that solving the non-linear program (NLP)

\begin{equation*}
    x^* = \arg \min_{x \in X} \phi(\Design,x),
\end{equation*}
with a global solver in every iteration does not present a better alternative. This also requires many evaluations and finding global solutions of arbitrary NLPs is quite difficult in general.

Other state-of-the-art approaches to find optimal experimental designs include reformulations of the optimization problem as a semi-definite program \cite{VanderbergheSemiDefinite,DuarteSDP,SagnolSDP} or a second-order cone program \cite{SagnolSOCP}. These programs can be solved very efficiently and are not based on an iterative algorithm. Additionally, these problems are convex and thus every local solutions is already globally optimal. However, the formulation as SDP (and SOCP) is again based on a grid $X_{grid}$ and we require the Fisher Information Matrices $\mu(x)$ at all grid points.

In this paper we motivate and introduce a novel, computationally efficient design of experiments algorithm which can be applied to high-dimensional models. This algorithm adaptively selects points to sample via an approximation of the directional derivative $\phi(\Design,x)$. Thus the algorithm does not require evaluations of the model $f$ on a fine grid. Using two examples from chemical engineering we illustrate that the novel algorithm can drastically reduce the runtime of the computations and relies on significantly less evaluations of the Jacobian $Df$. Hence, the new algorithm can be applied to models which were previously considered too complex to perform model-based design of experiment strategies.
 
\section{Results}\label{sec:Results}
In this section we introduce a novel design of experiments algorithm. We begin by giving a brief overview on the design of experiments theory. We focus on the main results which are used in our algorithm. We also introduce two existing design of experiments algorithms. Our new algorithm builds upon the ideas of these algorithms.


Next we give an introduction into the theory of Gaussian process regression. We use Gaussian process regression in our new algorithm to approximate the directional derivative of the design criterion.

We end the section by giving the new algorithm. We introduce two examples from chemical engineering and perform evaluations on these examples.

\subsection{Design of Experiments}\label{sec:Theory}

In design of experiments we consider a model $f$ mapping inputs from a design space $X$ to outputs in $Y$. The model depends on parameters $\theta \in \Theta$ and is given by

\begin{equation*}
    f \colon X \times \Theta \to Y, (x,\theta) \mapsto f(x,\theta).
\end{equation*}
We let the spaces $X, \Theta$ and $Y$ be sub-spaces of $\R^{d_X}, \R^{d_\theta}$ and $\R^{d_Y}$ respectively.

A \emph{continuous design} $\Design$ is a set of tuples $(x_i,w_i)$ for $i=1,\ldots,n$, where the $x_i \in X$ are design points and the weights $w_i$ are positive and sum to $1 = \sum_{i=1}^n w_i$. We often denote 

\begin{equation*}
    \Design = \left\{ \begin{array}{ccc}
       x_1  & \cdots & x_n  \\
        w_1  & \cdots & w_n 
    \end{array}\right\}.
\end{equation*}
The design points $x_i$ with $w_i>0$ are called the support points of the design. We can generalize the notion of a continuous design by considering \emph{design measures} $\Design$, which correspond to a probability measure on the design space $X$. 

For a design point $x$ we define the \emph{Fisher Information Matrix} $\mu(x)$ as 

\begin{equation*}
    \mu(x) = D_\theta f(x) \Sigma_{\varepsilon}^{-1} D_\theta f(x)^T,
\end{equation*}
where $D_\theta f(x)$ denotes the Jacobian matrix of $f$ with respect to $\theta$. For a continuous design $\Design$ the \emph{(normalized) Fisher Information Matrix} is given as weighted sum of the Fisher Information Matrices $\mu(x_i)$, 

\begin{equation*}
    \Fim(\Design) = \sum_{i=1}^n w_i \mu(x_i).
\end{equation*}
Note, that the Jacobian $D_\theta f(x)$ depends on the unknown parameters $\theta$ and therefore so does the Fisher Information Matrix. As stated in Section \ref{sec:Introduction}, we insert our best current estimate of the parameters $\theta$ and compute locally optimal designs dependent on this value.

In design of experiments we want to find a design $\Design$ which minimizes a measurement function of the Fisher Information Matrix, the \emph{design criterion}. Commonly used design criteria $\Criteria$ include the \begin{itemize}[leftmargin=*,labelsep=5.8mm]
    \item A-Criterion. This criterion corresponds to the trace of the inverse of the Fisher Information Matrix: 
    
    \begin{equation*}
        \Criteria_A (M(\DesignN)) = tr\left[  M(\DesignN)^{-1} \right].
    \end{equation*}
    \item D-Criterion. Here we consider the determinant of the inverse of the Fisher Information Matrix: 
    
    \begin{equation*}
        \Criteria_D(M(\DesignN)) = det\left[ M(\DesignN) ^{-1} \right] = det\left[  M(\DesignN)  \right]^{-1}.
    \end{equation*}
    Equivalently we often consider the logarithm of the determinant, which is then called the \emph{log-D-Criterion}.
    
    \item E-Criterion. This criterion is the largest eigenvalue $\lambda_{max}$ of the Fisher Information Matrix. Equivalently we can consider the smallest eigenvalue $\lambda_{min}$ of its inverse: 
    
    \begin{equation*}
        \Criteria_E (M(\DesignN)) = \lambda_{max}(M(\DesignN)) = \frac{1}{\lambda_{min}(M(\DesignN)^{-1})}.
    \end{equation*}
\end{itemize}

The design of experiments problem is then given as

\begin{equation*}
    \begin{aligned}
      \min_{n, w_i, x_i} &  \Criteria\left( \sum_{i=1}^n w_i \cdot \mu(x_i) \right) \\
     s.t.\quad & \sum_{i=1}^n w_i = 1, \ 0 \leq w_i, \\
     & x_i \in X, \ n \in \N .
    \end{aligned}
\end{equation*}

We also denote this problem by 

\begin{equation}\label{eq:Problem2}
    \min_{\Design \in \DESIGN{X}} \Criteria(M(\Design)),
\end{equation}
where $\DESIGN{X}$ corresponds to the space of all design measures on $X$.

Under mild assumptions on the design criterion $\Criteria$ and the design space $X$ optimality conditions for the Optimization Problem \ref{eq:Problem2} are known. We refer to \cite[Chapter 2]{FederovBook} for these assumptions. In particular we require the directional derivative $\phi(\Design,x)$ of the design criterion $\Criteria$. These derivates are known for the A-, E- and (log-)D-Criterion and given by \begin{itemize}[leftmargin=*,labelsep=5.8mm]
    \item $\phi_D(\Design,x) = d_\Theta - tr\left[\Fim(\Design)^{-1} \mu(x) \right] $, where $d_\Theta$ is the number of unknown model parameters.
    \item $\phi_A(\Design,x) = tr\left[\Fim(\Design)^{-1} - \Fim(\Design)^{-2} \mu(x) \right] $.
    \item $\phi_E(\Design,x) = \lambda_{min}(\Fim(\Design)) - \sum_{i=1}^{mult(\lambda_{min})} \pi_i P_i^T \mu(x) P_i$, where $mult(\lambda_{min})$ is the algebraic multiplicity of $\lambda_{min}$, the $\pi_i$ are positive factors summing to unity and the $P_i$ are normalized, linear independent eigenvectors of $\lambda_{min}$.
\end{itemize}

The following theorems give global optimality conditions and can be found in \cite{FederovBook,Atkinson}.

\begin{Theorem}\label{the:OptimalityConditions}
The following holds:
\begin{itemize}[leftmargin=*,labelsep=5.8mm]
    \item An optimal design $\Design^*$ exists, with at most $\frac{d_\Theta(d_\Theta + 1)}{2}$ support points.
    \item The set of optimal designs is convex.
    \item The condition
    
    \begin{equation*}
        \min_{x \in X} \phi(\Design^*,x) \geq 0.
    \end{equation*} is necessary and sufficient for the design $\Design^*$ to be (globally) optimal.
    \item For $\Design^*$ almost-every support point of $\Design^*$ we have $\phi(\Design^*,x) = 0$.
\end{itemize}
\end{Theorem}

Typically, the optimality conditions presented in Theorem \ref{the:OptimalityConditions} are known as the \emph{Equivalence Theorems}. These present a reformulation of the results from Theorem \ref{the:OptimalityConditions} and are contributed to \cite{KieferEquivalenceTheorem}.

\begin{Theorem}[Equivalence Theorem]\label{the:EquivalenceTheorem}
The following optimization problems are equivalent:
\begin{itemize}[leftmargin=*,labelsep=5.8mm]
    \item $\min_{\Design \in \DESIGN{X}} \Criteria(M(\Design))$
    \item $\max_{\Design \in \DESIGN{X}} \min_{x \in X} \phi(x,\Design)$
    \item $\min_{x \in X} \phi(x,\Design) = 0$.
\end{itemize}
\end{Theorem}

\begin{Remark}\label{rem:D_Crit}
Performing the experiments of a design $\Design$ we obtain an estimate $\theta_{est}$ of the model parameters $\theta$. With this estimate we can give a prediction on the model outputs $y_{pred} = f(x,\theta_{est})$. The variance of this prediction is given by

\begin{equation*}
    \Var{y_{pred}} = tr\left[\Fim(\Design)^{-1} \mu(x) \right].
\end{equation*}
For details on the computations we refer to \cite[Chapter 2.3]{FederovBook}

We observe, that the directional derivative $\phi_D$ of the D-Criterion is given by $\phi_D(\Design,x) = d_\Theta - \Var{y_{pred}}$. The Equivalence Theorem for the D-Criterion thus states that an optimal design $\Design^*$ minimizes the maximum variance of the prediction $y_{pred}$. As the variance is an indication on the error in the prediction, we also (heuristically) say that the design $\Design^*$ minimizes the maximum prediction error.
\end{Remark}

Based on the Kiefer-Wolfowitz Equivalence Theorem \ref{the:EquivalenceTheorem}, a variety of algorithms have been derived to compute optimal continuous designs. We introduce two such algorithms and begin with the \emph{Vertex Direction Method (VDM)}, which is sometimes also called \emph{Federov-Wynn Algorithm}. 

From Theorem \ref{the:OptimalityConditions} it follows that the support of an optimal design coincides with the minima of the function $\phi(\Design,x)$. Thus the minimum of $\phi(\Design,x)$ is of particular interest. We recall from Remark \ref{rem:D_Crit}, that this corresponds to the maximum prediction error for the D-Criterion. These considerations result in an iterative scheme where we 


\begin{itemize}[leftmargin=*,labelsep=5.8mm]
    \item compute the minimum $x^* = \arg \min_{x \in X} \phi(\Design,x)$ for a given design $\Design = \{ (x_1,w_1),\ldots,(x_n,w_n) \}$ and then
    \item add the point $x^*$ to the support of $\Design$.
\end{itemize}

Adding a support point to the design requires a redistribution of the weights. In the VDM weights are uniformly shifted from all previous support points $x_1, \ldots, x_n$ to the new support point $x^*$. We assign the point $x^*$ the weight $w^* = \alpha$ and set the remaining weights to $w_i \to (1-\alpha)\cdot w_i$ for an $\alpha \in [0,1]$. In \cite[Chapter 3]{FederovBook} and \cite{WynnAlgorithm} a detailed description of the algorithm with suitable choices of $\alpha$ and proof of convergence is given.

Next we state the \emph{YBT Algorithm}, introduced in \cite{YBTAlgorithmus}. This algorithm improves the distribution of the weights in each iteration and thus converges in less iterations to a (near) optimal design.

In the VDM we uniformly shift weights to the new support point $x^*$, in the YBT Algorithm we instead distribute the weights optimally among a set of candidate points. For a given set $\{x_1,\ldots,x_n\}$ we thus consider the optimization problem 

\begin{equation}\label{eq:WeightProblem}
    \begin{aligned}
      \min_{w_i}\ &  \Criteria\left( \sum_{i=1}^n w_i \cdot \mu(x_i) \right) \\
     s.t.\quad & \sum_{i=1}^n w_i = 1, \ 0 \leq w_i.
    \end{aligned}
\end{equation}

With an optimal solution $w^*$ of Problem \ref{eq:WeightProblem} we then obtain a design $\Design$ by assigning each design point $x_i$ the weight $w^*_i$.

For the YBT-Algorithm we solve the optimization Problem \ref{eq:WeightProblem} in each iteration which results in the following iterative scheme:

\begin{itemize}[leftmargin=*,labelsep=5.8mm]
    \item For the candidate point set $X_n = \{ x_1, \ldots, x_n\}$ solve Problem \ref{eq:WeightProblem} to obtain optimal weights $w^*$.
    \item Obtain the design $\Design_n$ by combining the candidate points $X_n$ with the optimal weights $w^*$. 
    \item Solve $x_{n+1} = \arg \min_{x \in X} \phi(\Design_n,x)$.
    \item If $\phi(\Design_n,x_{n+1})>-\varepsilon$, the design $\Design_n$ is (near) optimal.
    \item Else, add $x_{n+1}$ to the candidate points $X_n$ to obtain $X_{n+1} = \{ x_1, \ldots, x_{n+1} \}$ and iterate.
\end{itemize}
In the YBT Algorithm we solve two optimization problems in each iteration $n$. The weight optimization \ref{eq:WeightProblem} is a convex optimization problem \cite{boyd_vandenberghe_2004,Vanaret2020}. In \cite{YBTAlgorithmus} it is proposed to use an optimization based on Newton's Method. However, Problem \ref{eq:WeightProblem} can also be reformulated as a semidefinite program (SDP) or as second-order conic program (SOCP), see \cite{DuarteSDP,VanderbergheSemiDefinite,boyd_vandenberghe_2004,SagnolSDP} and \cite{SagnolSOCP}. Both SDPs and SOCPs can be solved very efficiently and we recommend reformulating the weight optimization problem \ref{eq:WeightProblem}.

The optimization of the directional derivative $\varphi(\Design,x)$ on the other hand is not convex in general. A global optimization is therefore difficult. A typical approach to resolve this issue is to consider a finite design space $X_{grid}$. We then evaluate the function $\varphi(\Design,x)$ for every $x \in X_{grid}$ to obtain the global minimum. For continuous design spaces $X$ we substitute the design space with a fine equidistant grid $X_{grid} \subset X$. This approach is also utilized by other state-of-the-art algorithms like Yu's Cocktail Algorithm \cite{YuCocktail} or the Random Exchange Method \cite{HarmanRandExchangeAlg}.

When considering a finite design space $X_{grid}$, we can also solve the weight optimization \ref{eq:WeightProblem} on the whole design space $X_{grid}$ to obtain an optimal design. Depending on the size of the design space $X_{grid}$ this results in a very large optimization problem with many optimization variables. An adaptive algorithm like the YBT Algorithm can drastically decrease the size of this optimization problem at the cost of solving the problem repeatedly.

\subsection{Gaussian Process Regression}\label{sec:GPR}

Gaussian process regression (GPR) is a machine learning method used to approximate functions. We use this method to approximate the directional derivative $\phi(\Design,x)$ in our novel DoE algorithm. For the D-Criterion, this corresponds to an approximation of the prediction error of the model, see Remark \ref{rem:D_Crit}. Here we give a brief introduction into the theory, then we comment on some considerations for the implementation. For details we refer to \cite[Chapter 2]{RasWilBook}.


We consider a function $g \colon X \to \R$ which we want to approximate by a function $\tilde{g}(x)$. As we have no information on the value $g(x)$, we assume these values to follow a prior distribution $P_{prior}$. We then evaluate the function $g$ on a set of inputs $X_t$. Next the distribution $P_{prior}$ is conditioned on the tuples $(x_i,g(x_i))$, for $x_i \in X_t$, to obtain a posterior distribution $P_{post}$. This posterior distribution allows us to make more reliable predictions on the values $g(x)$.

In Gaussian process regression we assume the prior distribution to be given via a Gaussian process $G$ and the values to be normal distributed $g(x) \sim N(\mu_x, \sigma_x^2)$. The process $G$ is defined by its mean $m \colon X \to \R$ and its covariance kernel $k \colon X \times X \to \R$. The covariance kernel hereby always is a symmetric non-negative definite function. In the following we take the mean $m$ as zero - $m(x)=0$ for all $x \in X$ - as is usual in GPR \cite{RasWilBook}.

For any point set $X_t = \{x_1,\ldots,x_n \}$ the prior distribution of $(g(x_1), \ldots, g(x_n))^T$ is a multivariate normal distribution with mean $m(X_t)$ and covariance matrix $k(X_t,X_t)$. Here we denote the vector $(m(x_1),\ldots,m(x_n))^T$ by $m(X_t)$ and the matrix $(k(x_i,x_j))_{i,j=1,\ldots,n}$ by $k(X_t,X_t)$.  

We now assume to have access to evaluations of $g$ on the input set $X_t$, meaning that the vector $g(X_t)$ is known. We condition the random variable $g(x)$ on these values and denote the conditioned variable by $g(x)\left| X_t, g(X_t) \right.$. The distribution of $g(x)\left| X_t, g(X_t) \right.$ is again a normal distribution with mean 

\begin{equation}\label{eq:PostExpectation}
    \E{g(x)\left| X_t, g(X_t)\right.} = k(x,X_t) k(X_t,X_t)^{-1} g(X_t)
\end{equation}
and variance 

\begin{equation}\label{eq:PostVariance}
    \Var{g(x)\left| X_t, g(X_t)\right.} = k(x,x) - k(x,X_t) k(X_t,X_t)^{-1} k(X_t,x),
\end{equation}
where we use the notation $k(x,X_t) = (k(x,x_1), \ldots, k(x,x_n))$ and $k(X_t,x) = k(x,X_t)^T$. These values can be computed using Bayes Rule, a detailed description is given in \cite{RasWilBook}. The data pair $X_t, g(X_t)$ is called the \emph{training data} of the GPR.

The posterior expectation given in Equation \ref{eq:PostExpectation} is now used to approximate the unknown function $g$. We therefore set our approximation as $\tilde{g}(x) \coloneqq \E{g(x)\left| X_t, g(X_t)\right.}$. The posterior variance given in Equation \ref{eq:PostVariance} on the other hand is an indicator of the quality of the approximation. We note in particular that the approximation $\tilde{g}(x)$ interpolates the given training data $X_t, g(X_t)$, such that for every $x_i \in X_t$ we have

\begin{equation*}
    \tilde{g}(x_i) = \E{g(x_i)\left| X_t, g(X_t)\right.} = g(x_i).
\end{equation*}
For the training points $x_i \in X_t$ it holds that $\Var{g(x_i)\left| X_t, g(X_t)\right.} = 0$, also indicating that the approximation is exact.

The approximating function $\tilde{g}(x)$ also is a linear combination of the kernel functions $k(x,x_i)$. Via the choice of the kernel $k$, we can thus assign properties to the function $\tilde{g}(x)$. One widely used kernel is the squared exponential kernel $k_{sq}$ given by

\begin{equation*}
    k_{sq}(x,y) = \exp\left( - \frac{\| x-y \|_2^2}{2} \right).
\end{equation*}
For this kernel the approximation $\tilde{g}(x)$ always is a $C^\infty$ function. Other popular kernels are given by the \emph{Matern} class of kernels $k_\nu$, which are functions in $C^\nu$. The covariance kernel $k$ - in contrast to the mean $m$ - thus has a large influence on the approximation.

We now comment on two details of the implementation of Gaussian process regression. First we discuss the addition of White noise to the covariance kernel $k$. Here an additional White noise term is added to the kernel to obtain 

\begin{equation*}
    \tilde{k}(x,y) = k(x,y) + \sigma_W^2 \cdot \delta(x,y),
\end{equation*}
where $\delta(x,y)$ denotes the delta function. This function takes the values $\delta(x,x) = 1$ and $\delta(x,y) = 0$ if $x \neq y$. For the new kernel $\tilde{k}$ with $\sigma_W^2>0$ the matrix $\tilde{k}(X_t,X_t)$ always is invertible. However, the approximation $\tilde{g}$ arising from the adapted kernel $\tilde{k}$ need not interpolate the data and instead allows for deviations $\tilde{g}(x_i) = g(x_i) + \varepsilon$. The variance of these deviations is given by $\sigma_W^2$.

Second, we discuss the hyper-parameter selection of the kernel $k$. Often the kernel $k$ depends on additional hyper-parameters $\sigma$. For the squared exponential kernel 

\begin{equation*}
    k_{sq}(x,y) = \sigma_f^2 \cdot \exp\left( - \frac{\| x-y \|_2^2}{2l^2} \right) + \sigma_W^2 \cdot \delta(x,y)
\end{equation*}
these parameters are given as $\sigma = (\sigma_f^2,\sigma_W^2,l)$.

In order to set appropriate values of $\sigma$ we consider a loss function $L(X_t,g(X_t),k,\sigma) \in \R_{\geq 0}$. The hyper-parameters $\sigma$ are then set to minimize the loss $L(X_t,g(X_t),k,\sigma)$. Examples and discussions of loss functions can be found in \cite[Chaper 4]{RasWilBook}. For our implementation we refer to Section \ref{sec:Materials}.

\subsection{Novel Design of Experiments Algorithm}\label{sec:ADAGPR}

We now introduce our novel design of experiments algorithm, the \emph{ADA-GPR}. This algorithm builds upon the state-of-the-art YBT Algorithm and relies upon Gaussian process regression.

As previously noted, the YBT Algorithm is typically applied to a fine grid $X_{grid}$ in the continuous design space $X$. This adjustment is made, as finding a global optimum of the directional derivative $\phi(\Design,x)$ is challenging in general. For each point of the design space $X_{grid}$ we have to evaluate the Jacobian $D_\theta f(x)$ in order to compute the Fisher Information Matrix $\mu(x)$. We thus have to pre-compute the Fisher Information Matrices at every point $x \in X_{grid}$ for grid-based DoE algorithms.

In order to obtain reliable designs the grid $X_{grid}$ has to be a fine grid in the continuous design space $X$. The number of points increases exponentially in the dimension and so then does the pre-computational time required to evaluate the Jacobians. Grid-based methods like the YBT Algorithm are therefore problematic when we have a high-dimensional design space and can lead to very long runtimes. For particular challenging models, they may not be viable at all.

The aim of our novel algorithm is to reduce the evaluations of the Fisher Information Matrices and thereby reduce the computational time for high-dimensional models.

We observe, that in the YBT Algorithm solely the Fisher Information Matrices of the candidate points $x_i \in X_n$ are used to compute optimal weights $w^*$. The matrices $\mu(x)$ at the remaining points are only required to find the minimum $x^* = \arg \min \phi(\Design_n,x)$.

In order to reduce the number of evaluations, we thus propose to only evaluate the Jacobians and the Fisher Information Matrices of the candidate points $X_n$. With the Jacobians at these points we can compute exact weights $w^*$ for the candidate points. The directional derivative $\phi(\Design_n,x)$ however is approximated in each iteration of the algorithm. For the approximation we use Gaussian process regression. As training data for the approximation we also use the candidate points $X_n$ and the directional derivative $\phi(\Design_n,x)$ at these points. As we have evaluated the Jacobians at these points, we can compute the directional derivative via matrix multiplication.

We briefly discuss why GPR is a viable choice for the approximation of $\phi(\Design_n,x)$. In the YBT Algorithm we iteratively increase the number of candidate points $X_n$. Thus we want an approximation which can be computed for an arbitrary amount of evaluations and which has a consistent feature set. As stated in Section \ref{sec:GPR} we can compute a GPR with arbitrary training points and we can control the features of the approximation via the choice of kernel $k$. Additionally, we not only obtain an approximation via GPR, but also the variance $\Var{\phi(\Design_n,x) \left| X_n,\phi(\Design_n,X_n) \right.}$. The variance gives information on the quality of the approximation, which can also be useful in our considerations.

In Section \ref{sec:GPR} we have discussed, that $\E{\phi(\Design_n,x)\left| X_n, \phi(\Design_n,X_n) \right.}$ is the appropriate choice to approximate the directional derivative $\phi(\Design_n,x)$. This suggests to select the upcoming candidate points via 

\begin{equation*}
    x_{n+1} = \arg \min_{x \in X} \E{\phi(\Design_n,x)\left| X_n, \phi(\Design_n,X_n) \right.}.
\end{equation*}

However, we also want to incorporate the uncertainty of the approximation into our selection. Inspired by Bayesian Optimization \cite{frazier2018tutorial,snoek2012practical,BayesOptLoop} we consider an approach similar to the Upper-Confidence Bounds (UCB). Here we additionally subtract the variance from the expectation for our point selection. This results in the following point

\begin{equation}\label{eq:SelectionFunction}
    x_{n+1} = \arg \min_{x \in X} \E{\phi(\Design_n,x)\left| X_n, \phi(\Design_n,X_n) \right.} - \Var{\phi(\Design_n,x)\left| X_n, \phi(\Design_n,X_n) \right.}.
\end{equation}
We call the function $\E{\ } - \Var{\ }$ used to select the next candidate point $x_{n+1}$ the \emph{acquisition function} of the algorithm. This denotation is inspired by Bayesian Optimization, too \cite{BayesOptLoop}.

The two terms $\E{\phi(\Design_n,x)\left| X_n, \phi(\Design_n,X_n) \right.}$ and $\Var{\phi(\Design_n,x)\left| X_n, \phi(\Design_n,X_n) \right.}$ hereby each represent an own objective. Optimization of the expectation $\E{\ }$ results in points which we predict to be minimizers of $\phi(\Design_n,x)$. These points then help improve our design and thereby the objective value. Optimization of the variance on the other hand leads to points where the current approximation may have large errors. Evaluating at those points improves the approximation in the following iterations. By considering the sum of both terms we want to balance these two goals. Additionally, we can introduce a parameter $\tau$ as a factor to the variance. We can then control how to weight the terms. However, we have found that $\tau = 1$ as is in Equation \ref{eq:SelectionFunction} gives good results.

We make one last adjustment to the point acquisition. We want to avoid having a bad approximation which does not correctly represent minima of the directional derivative. Thus - if the directional derivative at the new candidate point $x_{n+1}$ is not negative and $ \phi(\Design_n,x_{n+1}) >0  $ - we select the upcoming point only to improve the approximation. This is achieved by selecting the point $x_{n+2}$ according to 

\begin{equation*}
    x_{n+2} = \arg \max_{x \in X} \Var{\phi(\Design_{n+1},x)\left| X_n, \phi(\Design_{n+1},X_{n+1}) \right.}.
\end{equation*}
The uncertainty in the approximation is represented by the variance and evaluating at a point of high variance therefore increases the accuracy of the approximation. However, we do not use this selection in successive iterations.

For Gaussian process regression the derivatives of the expectation $\E{\ }$ and variance $\Var{\ }$ are also known. These can be given to the solver we use to compute $x_{n+1}$ and improve the performance. This is another advantage of the proposed approximation via GPR.

The proposed algorithm is given in Algorithm \ref{alg: ADAlgorithm}. We call this adaptive algorithm the \emph{ADA-GPR}.

    

\begin{algorithm}[H]
\caption{Adaptive Discretization Algorithm with Gaussian Process Regression}
\label{alg: ADAlgorithm}

\vspace{0.2cm}
Select an arbitrary initial candidate point set $X_{n_0} = \left\{ x_1, \ldots, x_{n_0} \right\} \subset X$.

\vspace{0.1cm}
Evaluate the Jacobians $D_\theta f(x_i)$ at the candidate points $x_i \in X_{n_0}$ and assemble these in the set $J_{n_0} = \left\{ D_\theta f(x_i) , x \in X_{n_0}\right\}$.

\vspace{0.1cm}
Set $\tau = 1$.

\vspace{0.1cm}
Iterate over $n \in \N$:
\begin{enumerate}[leftmargin=*,labelsep=4.9mm]
    \item Compute optimal weights $w^*$ for the candidate points and combine the candidate points $X_n$ and the weights $w^*$ to obtain the design $\Design_n$.
    
    \vspace{0.1cm}
    \item Compute the directional derivative at the candidate points $x_i \in X_n$ via the values in $D_\theta f(x_i) \in J_n$. Gather these in the set $Y_n = \left\{ \phi(\Design_n,x_i) , x_i \in X_n \right\}$.
    
    \vspace{0.1cm}
    \item Compute a GPR for the training data $\left( X_n, Y_n \right)$.
    
    \vspace{0.1cm}
    \item Solve $x_{n+1} = \arg \max_{x \in X}  \Var{\phi(\Design_n,x)\left|X_n,Y_n \right.} + \tau \cdot \E{\phi(\Design_n,x)\left|X_n,Y_n \right.} $.
    
    \vspace{0.1cm}
    \item Compute the Jacobian $D_\theta f(x_{n+1})$ and the directional derivative $\phi(\Design_n,x_{n+1})$ at the point $x_{n+1}$.
    
    \vspace{0.1cm}
    \item Update the sets $X_{n+1} = X_n \cup \{x_{n+1}\}$ and $J_{n+1} = J_n \cup \left\{ D_\theta f(x_{n+1}) \right\}$.
    
    \vspace{0.1cm}
    \item If $\varphi(\Design_n,x_{n+1}) < 0$:\quad Set $\tau = 1$.
    
    \vspace{0.1cm}
    \item Else if $\varphi(\Design_n,x_{n+1}) \geq 0$:
    \begin{itemize}
    \vspace{-0.2em}
        \item  If $\tau = 1$, set $\tau = 0$ for the next iteration. 
        
        \vspace{0.5em}
         \item If $\tau = 0$, set $\tau = 1$.
    \end{itemize}
\end{enumerate}
\end{algorithm}

\begin{figure}[H]
\centering
\includegraphics[width=0.45 \linewidth]{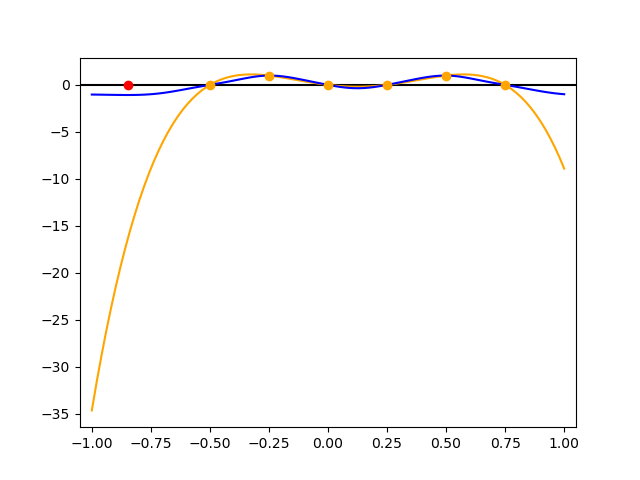}
\includegraphics[width=0.45 \linewidth]{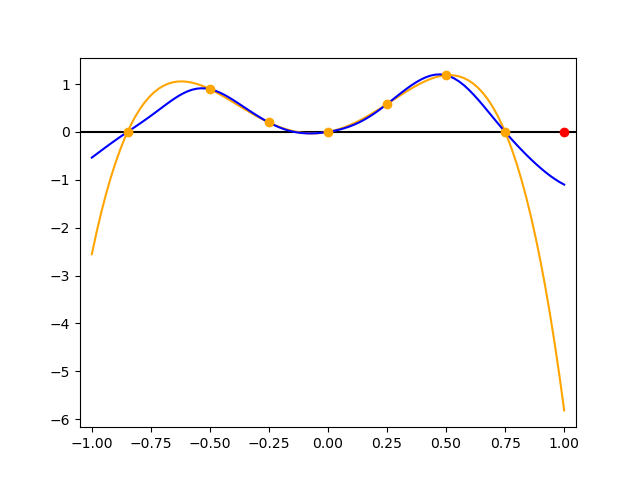}
\includegraphics[width=0.45 \linewidth]{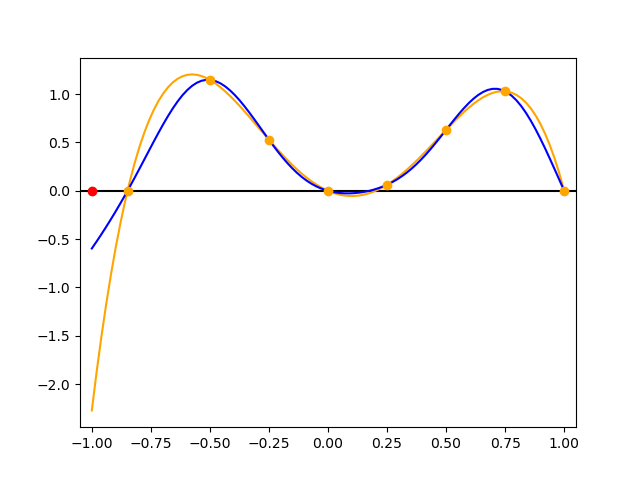}
\includegraphics[width=0.45 \linewidth]{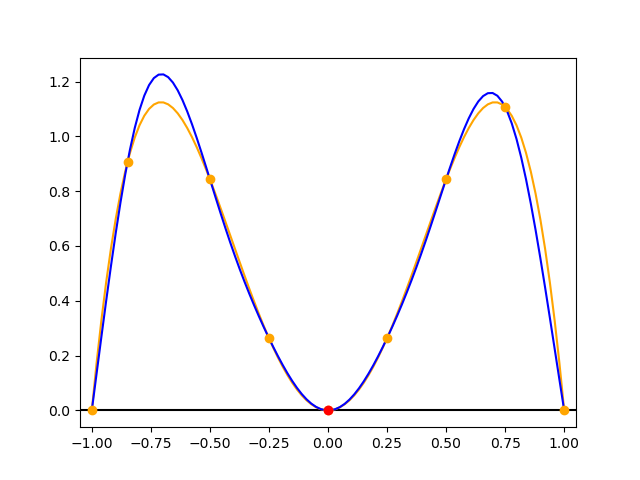}
\caption{Illustration of the point acquisition in the first $4$ iterations of the novel ADA-GPR with the D-Criterion. The directional derivative $\phi_D$ (orange), the acquisition function (blue) as well as the current candidate points $X_n$ (orange) and the new selected point $x_{n+1}$ (red) are plotted.}
\label{fig:Quad_Example}
\end{figure}

In Figure \ref{fig:Quad_Example} the adaptive point acquisition in the first $4$ iterations of the ADA-GPR for a quadratic toy example $f(x,\theta) = \theta_2 x^2 + \theta_1 x+ \theta$ is illustrated. We can observe how the algorithm selects the next candidate point $x_{n+1}$ in these illustrations. We note, that the point selected via the acquisition function can differ from the point we select via the exact values of $\phi_D$.

To end this section we discuss the quality of the design we obtain with the novel algorithm. For the YBT-Algorithm as well as the VDM, it is shown in \cite{YBTAlgorithmus,FederovBook} that the objective values converge to the optimal objective value with increasing number of iterations. We also have a stopping criterion which indicates, how far our current design is from being optimal \cite[Chapter 2]{FederovBook}. As we are using an approximation in the novel ADA-GPR, we cannot derive such results. The maximum error in our approximation will always present a bound for the error in the objective values of the design. However, we highlight once more that the aim of the ADA-GPR is to obtain an approximation of an optimal design with significantly less evaluations of the Jacobian $D_\theta f(x)$. For models which we cannot evaluate on a fine grid in the design space the ADA-GPR then presents a viable option to obtain designs in the continuous design space $X$.

\subsection{Chemical Engineering Examples}\label{sec:Examples}

We now provide two examples to illustrate the performance of the new algorithm compared to known ones. We evaluate the VDM, the YBT Algorithm (both Section \ref{sec:Theory}) and the new \emph{ADA-GPR} (Section \ref{sec:ADAGPR}). In this section we restrict ourselves to the log-D-Criterion for all algorithms. We then compare the results and the runtimes of the different methods. The models we present in this section are evaluated in CHEMASIM and CHEMADIS, the BASF in-house programs (Version $6.6$ \cite{ChemasimAspiron}), using the standard settings.

\begin{figure}[H]
\centering
\includegraphics[width=0.6 \linewidth]{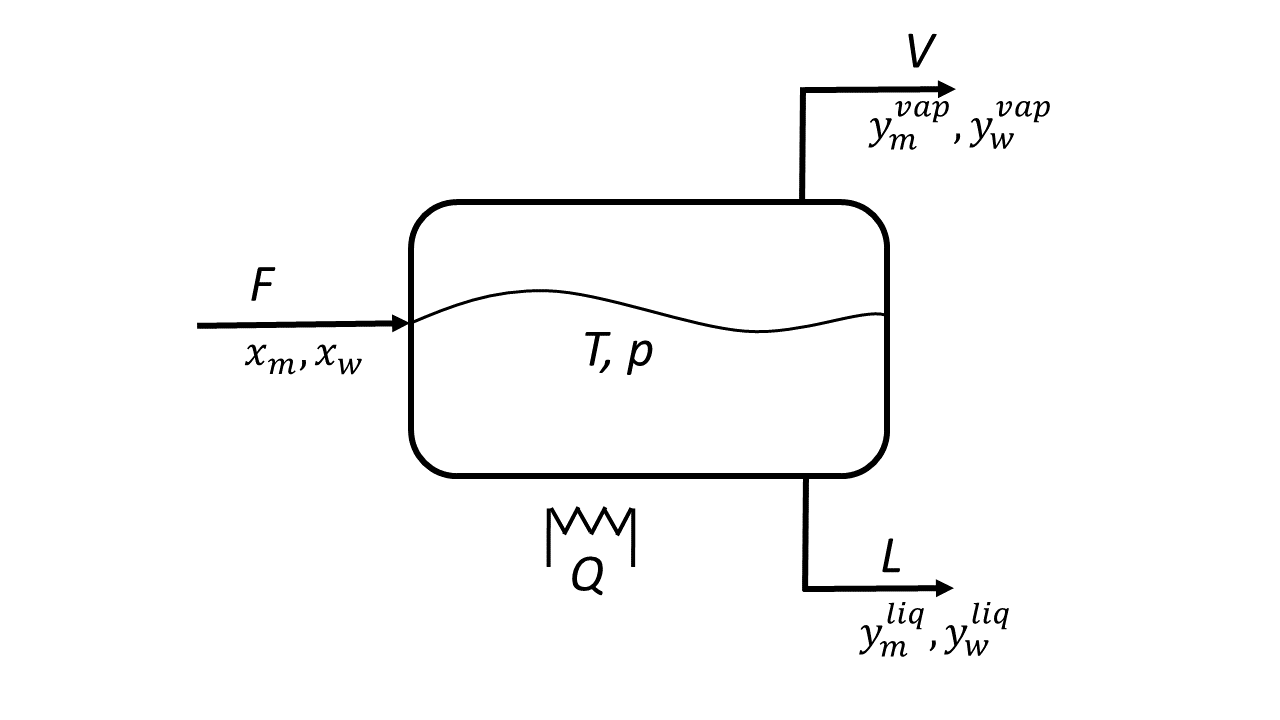}
\caption{Scheme of the flash.}
\label{fig:FlashUnit}
\end{figure}

The first example we consider is a flash. A liquid mixture consisting of two components enters the flash, where the mixture is heated and partially evaporates. One obtains an vapor liquid equilibrium at temperature $T$ and pressure $P$. In Figure \ref{fig:FlashUnit} the flash unit with input and output streams is sketched. The relations in the flash unit are governed by the \emph{MESH} equations (see Appendix \ref{app:MESH} and \cite{biegler1997}). 


Initially we consider a mixture composed of methanol and water. In a second step we replace the water component with acetone. The model and in particular the \emph{MESH} equations remain the same, however, parameters of the replaced component can vary. Details on the parameters for vapor pressure are given in the Appendix \ref{app:MESH}. 

In the following we denote the molar concentrations of the input stream by $x_m$ and $x_w$, those of the liquid output stream by $y_m^{liq}$ and $y_w^{liq}$ and those of the vapor output stream by $y_m^{vap}$ and $y_w^{vap}$. Additionally we introduce the molar flow rates $F$ of the input stream, $V$ of the vapor output stream and $L$ of the liquid output stream. 

We consider the following design of experiments setup:
\begin{itemize}[leftmargin=*,labelsep=5.8mm]
    \item two inputs, $P$ and $x_m$. Here $P$ denotes the pressure in the flash unit which ranges from $0.5$ to $5$ bar. The variable $x_m$ gives the molar concentration of methanol in the liquid input stream. This concentration is between $0$ and $1$ \nicefrac{mol}{mol}.
    \item two outputs, $T$ and $y^{vap}_m$. The temperature in degree Celsius at equilibrium $T$ is measured as well as the molar concentration $y^{vap}_m$ in \nicefrac{mol}{mol} of the evaporated methanol at equilibrium.
    \item four model parameters $a_{12}, a_{21}, b_{12}$ and $b_{21}$. These are parameters of the activity coefficients $\gamma_w$ and $\gamma_m$ in the \emph{MESH} equations (Appendix \ref{app:MESH}) - the so-called NRTL parameters.
\end{itemize}

We fix the flow rates $F = 1$ \nicefrac{kmol}{h} and $V = 10^{-6}$  \nicefrac{kmol}{h}. Given the inputs $P$ and $x_m$ we can then solve the system of equations to obtain the values of $T$ and $y^{vap}_m$. We represent this via the model function $f$ as

\begin{equation*}
    \left( \begin{array}{c}
         y_m^{vap} \\
         T
    \end{array} \right) = f\left(\left( x_m, P \right)^T,\left( a_{12}, a_{21}, b_{12}, b_{21}\right)^T  \right),
\end{equation*}
with $f\colon \left([0,1] \times [0.5,5]\right) \times \R^4 \ \to \ \R^2$.

For the VDM and the YBT Algorithm we need to replace the continuous design space with a grid. Therefore we set

\begin{equation*}
    X_{grid} = \left\{\left. \left( \frac{i}{100}\ \frac{\text{mol}}{\text{mol}}, \frac{10+j}{20}\ \text{bar}\ \right) \right| \ i=0, \ldots, 100,\ j = 0, \ldots, 90 \right\}.
\end{equation*}
This set corresponds to a grid with $9191$ design points. 

\begin{figure}[H]
\centering
\includegraphics[width=0.45 \linewidth]{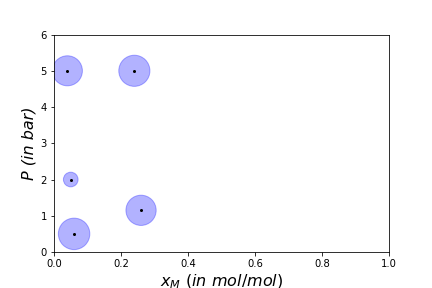}
\includegraphics[width=0.45 \linewidth]{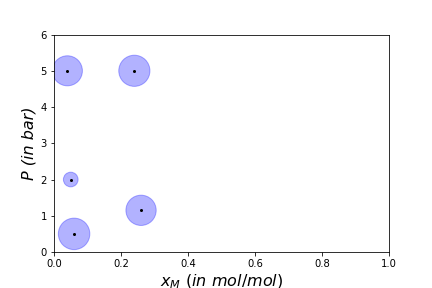}
\caption{Optimal designs in the two-dimensional design space for the flash with a methanol-water input feed. (\textbf{a}) The design $\Design_{VDM}$ computed with the VDM. (\textbf{b}) The design $\Design_{YBT}$ computed with the YBT Algorithm.}
\label{fig:DesignsMeOHWYBT}
\end{figure}

The designs $\Design_{VDM}$ and $\Design_{YBT}$ the algorithms compute are given in Table \ref{tab:MeOHWVDM} and in Figure \ref{fig:DesignsMeOHWYBT}. We note that only design points with a weight larger than $0.001$ are given - in particular for the VDM, where the weights of the initial points only slowly converge to $0$.
 
\begin{table}[H]
\caption{Optimal designs (\textbf{a}) $\Design_{VDM}$ computed with the Vector Direction Method, (\textbf{b}) $\Design_{YBT}$ computed with the YBT Algorithm and (\textbf{c}) $\Design_{ADAGPR}$ computed with the ADA-GPR for the flash with a methanol-water input feed.}
\label{tab:MeOHWVDM}
\centering
\begin{tabular}[t]{cccc}
\toprule
\textbf{}	& \textbf{$x_M$}	& \textbf{$P$} & \textbf{Weight}\\
\midrule
1 & 0.06 \nicefrac{mol}{mol} & 0.50 bar & 0.2477  \\
2 & 0.05 \nicefrac{mol}{mol} & 2.00 bar & 0.0538 \\ 
3 & 0.04 \nicefrac{mol}{mol} & 5.00 bar & 0.2258 \\ 
4 & 0.24 \nicefrac{mol}{mol} & 5.00 bar & 0.2426 \\
5 & 0.26 \nicefrac{mol}{mol} & 1.15 bar & 0.2287 \\ 
\bottomrule
\end{tabular}
\quad
\begin{tabular}[t]{cccc}
\toprule
\textbf{}	& \textbf{$x_M$}	& \textbf{$P$} & \textbf{Weight}\\
\midrule
1 & 0.04 \nicefrac{mol}{mol} & 5.00 bar & 0.2259 \\ 
2 & 0.06 \nicefrac{mol}{mol} & 0.50 bar & 0.2480 \\ 
3 & 0.05 \nicefrac{mol}{mol} & 2.00 bar & 0.0539 \\ 
4 & 0.24 \nicefrac{mol}{mol} & 5.00 bar & 0.2430 \\ 
5 & 0.26 \nicefrac{mol}{mol} & 1.15 bar & 0.2292 \\ 
\bottomrule
\end{tabular}

\quad

\begin{tabular}[t]{cccc}
\toprule
\textbf{}	& \textbf{$x_M$}	& \textbf{$P$} & \textbf{Weight}\\
\midrule
1 & 0.0667 \nicefrac{mol}{mol} & 0.5000 bar & 0.2487 \\
2 & 0.0415 \nicefrac{mol}{mol} & 4.6000 bar & 0.2367 \\ 
3 & 0.2642 \nicefrac{mol}{mol} & 1.1209 bar & 0.2402 \\ 
4 & 0.2401 \nicefrac{mol}{mol} & 5.0000 bar & 0.2464 \\ 
5 & 0.0492 \nicefrac{mol}{mol} & 1.8857 bar & 0.0271 \\
\bottomrule
\end{tabular}
\end{table}

\begin{figure}[H]
\centering
\includegraphics[width=0.45 \linewidth]{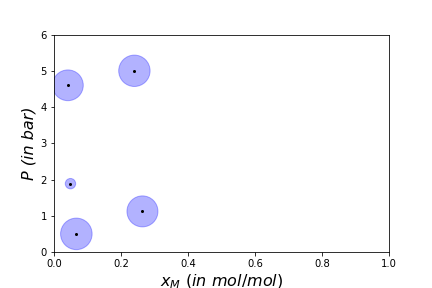}
\caption{Optimal design $\Design_{ADAGPR}$ in the two-dimensional design space for the flash with a methanol-water input feed, computed with the novel ADA-GPR.}
\label{fig:DesignsMeOHWADAGPR}
\end{figure}

The ADA-GPR on the other hand is initialized with $50$ starting candidate points. We obtain the design $\Design_{ADAGPR}$ given in Table \ref{tab:MeOHWVDM} and in Figure \ref{fig:DesignsMeOHWADAGPR}. Here we also only give points with a weight larger than $0.001$. Additionally, we have clustered some of the design points. We refer to Section \ref{sec:Materials} for details on the clustering.


In Table \ref{tab:MeOHWObjective} the objective value, the total runtime, the number of iterations and the number of evaluated Jacobians $Df$ is listed. The objective value is the logarithm in base $10$ of the determinant of the Fisher Information Matrix: $\log_{10} \left( \det \left(\Fim \right)\right)$. This corresponds to the negative D-Criterion and thus we aim at maximizing this value. A detailed breakdown of the runtime is given in Table \ref{tab:MeOHWRuntime}. Here we differentiate between the time contributed to evaluations of the Jacobians $D_\theta f$, the optimization of the weights $w_i$, the optimization of the hyper-parameters $\sigma$ of the GPR and the optimization of the point acquisition function.

\begin{table}[H]
\caption{Objective values, runtime and number of iterations of all three algorithms for the flash with a methanol-water input feed.}
\label{tab:MeOHWObjective}
\centering
\begin{tabular}{cccc}
\toprule
\textbf{}	& \textbf{VDM}	& \textbf{YBT Algorithm} & \textbf{ADA-GPR}\\
\midrule
Objective value & $7.9327$  & $7.9334$ & $7.9124$ \\
Iterations & $10000$ & $10$ & $136$ \\ 
Evaluations of the Jacobian & $9191$ & $9191$ & $151$ \\ 
Runtime & $4241.88$ s & $1790.21$ s & $124.10$ s \\ 
\bottomrule
\end{tabular}
\end{table}

\begin{table}[H]
\caption{Detailed breakdown of the runtime of all algorithms for the flash with a methanol-water input feed.}
\label{tab:MeOHWRuntime}
\centering
\begin{tabular}{cccc}
\toprule
\textbf{Runtime}	& \textbf{VDM}	& \textbf{YBT Algorithm} & \textbf{ADA-GPR}\\
\midrule
Total & $4241.88$ s & $1790.21$ s & $124.10$ s \\
Jacobian evaluation &  $1787.22$ s &$1787.22$ s &$25.20$ s \\
Optimization: weights & - &$0.34$ s &$4.41$ s\\ 
Optimization: acquisition function & $2447.71$ s &$2.52$ s &$46.44$ s\\ 
Optimization: hyper-parameters & - & - & $43.01$ s \\ 
\bottomrule
\end{tabular}
\end{table}

\begin{figure}[H]
\centering
\includegraphics[width=0.45 \linewidth]{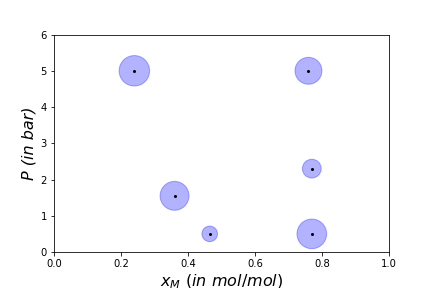}
\includegraphics[width=0.45 \linewidth]{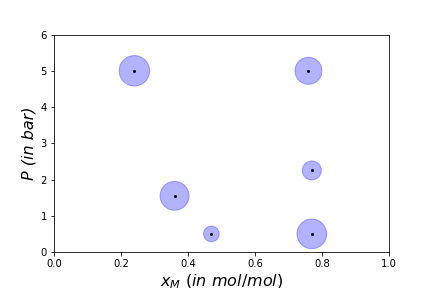}
\caption{Optimal designs in the two-dimensional design space for the flash with a methanol-acetone input feed. (\textbf{a}) The design $\Design_{VDM}$ computed with the VDM. (\textbf{b}) The design $\Design_{YBT}$ computed with the YBT Algorithm.}
\label{fig:DesignsMeOHAceYBT}
\end{figure}

\begin{figure}[H]
\centering
\includegraphics[width=0.45 \linewidth]{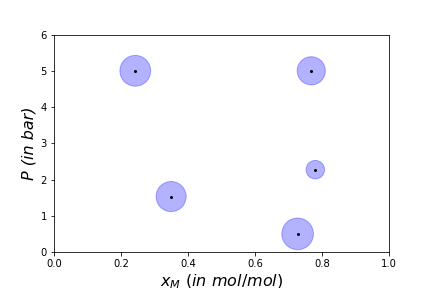}
\caption{Optimal design $\Design_{ADAGPR}$ in the two-dimensional design space for the flash with a methanol-acetone input feed computed with the novel ADA-GPR.}
\label{fig:DesignsMeOHAceADAGPR}
\end{figure}

Next we replace the mixture of water and methanol by a new mixture consisting of methanol and acetone. The designs $\Design_{VDM}, \Design_{YBT}$ and $\Design_{ADAGPR}$ - computed by the VDM, YBT Algorithm and the ADA-GPR respectively - are given in Table \ref{tab:MeOHAceVDM} as well as in Figures \ref{fig:DesignsMeOHAceYBT} and \ref{fig:DesignsMeOHAceADAGPR}. We initialize the algorithms with the same number of points as for the water-methanol mixture. A detailed breakdown of the objective values, number of iterations and the runtime is given in Tables \ref{tab:MeOHAceObjective} and \ref{tab:MeOHAceRuntime}.

\begin{table}[H]
\caption{Optimal designs (\textbf{a}) $\Design_{VDM}$ computed with the Vector Direction Method, (\textbf{b}) $\Design_{YBT}$ computed with the YBT Algorithm and (\textbf{c}) $\Design_{ADAGPR}$ computed with the ADA-GPR for the flash with a methanol-acetone input feed.}
\label{tab:MeOHAceVDM}
\centering
\begin{tabular}[t]{cccc}
\toprule
\textbf{}	& \textbf{$x_M$}	& \textbf{$P$} & \textbf{Weight}\\
\midrule
1 & 0.76 \nicefrac{mol}{mol} & 5.00 bar & 0.1816  \\
2 & 0.24 \nicefrac{mol}{mol} & 5.00 bar & 0.2324 \\ 
3 & 0.36 \nicefrac{mol}{mol} & 1.55 bar & 0.2092 \\ 
4 & 0.77 \nicefrac{mol}{mol} & 0.50 bar & 0.2199 \\
5 & 0.47 \nicefrac{mol}{mol} & 0.50 bar & 0.0605 \\ 
6 & 0.77 \nicefrac{mol}{mol} & 2.30 bar & 0.0887 \\ 
\bottomrule
\end{tabular}
\quad
\begin{tabular}[t]{cccc}
\toprule
\textbf{}	& \textbf{$x_M$}	& \textbf{$P$} & \textbf{Weight}\\
\midrule
1 & 0.24 \nicefrac{mol}{mol} & 5.00 bar & 0.2328 \\ 
2 & 0.77 \nicefrac{mol}{mol} & 0.50 bar & 0.2210 \\ 
3 & 0.47 \nicefrac{mol}{mol} & 0.50 bar & 0.0613 \\ 
4 & 0.36 \nicefrac{mol}{mol} & 1.55 bar & 0.2096 \\ 
5 & 0.76 \nicefrac{mol}{mol} & 5.00 bar & 0.1831 \\ 
6 & 0.77 \nicefrac{mol}{mol} & 2.25 bar & 0.0914 \\ 
\bottomrule
\end{tabular}

\quad

\begin{tabular}[t]{cccc}
\toprule
\textbf{}	& \textbf{$x_M$}	& \textbf{$P$} & \textbf{Weight}\\
\midrule
1 & 0.7277 \nicefrac{mol}{mol} & 0.5000 bar & 0.2499 \\
2 & 0.7684 \nicefrac{mol}{mol} & 5.0000 bar & 0.1982 \\ 
3 & 0.2428 \nicefrac{mol}{mol} & 5.0000 bar & 0.2369 \\ 
4 & 0.3498 \nicefrac{mol}{mol} & 1.5298 bar & 0.2259 \\ 
5 & 0.7805 \nicefrac{mol}{mol} & 2.2716 bar & 0.0860 \\
\bottomrule
\end{tabular}
\end{table}

\begin{table}[H]
\caption{Objective values, runtime and number of iterations of all three algorithms for the flash with a methanol-acetone input feed.}
\label{tab:MeOHAceObjective}
\centering
\begin{tabular}{cccc}
\toprule
\textbf{}	& \textbf{VDM}	& \textbf{YBT Algorithm} & \textbf{ADA-GPR}\\
\midrule
Objective value & $18.5060$  & $18.5064$ & $18.5020$ \\
Iterations & $10000$ & $16$ & $58$ \\ 
Evaluations of the Jacobian & $9191$ & $9191$ & $77$ \\ 
Runtime & $4287.86$ s & $1840.21$ s & $56.17$ s \\ 
\bottomrule
\end{tabular}
\end{table}

\begin{table}[H]
\caption{Detailed breakdown of the runtime of all algorithms for the flash with a methanol-acetone input feed.}
\label{tab:MeOHAceRuntime}
\centering
\begin{tabular}{cccc}
\toprule
\textbf{Runtime}	& \textbf{VDM}	& \textbf{YBT Algorithm} & \textbf{ADA-GPR}\\
\midrule
Total & $4287.86$ s & $1840.21$ s & $56.17$ s \\
Jacobian evaluation &  $1835.99$ s & $1835.99$ s &$13.69$ s \\
Optimization: weights & - &$0.18$ s &$0.80$ s\\ 
Optimization: acquisition function & $2443.89$ s &$3.86$ s &$16.55$ s\\ 
Optimization: hyper-parameters & - & - & $24.08$ s \\ 
\bottomrule
\end{tabular}
\end{table}

The second example we consider is the fermentation of baker's yeast. This model is taken from \cite{AspreyRobust,BarzUncertainty}, where a description of the model is given and DoE results for uncertain model parameters $\theta$ are presented.

\begin{figure}[H]
\centering
\includegraphics[width=0.8 \linewidth]{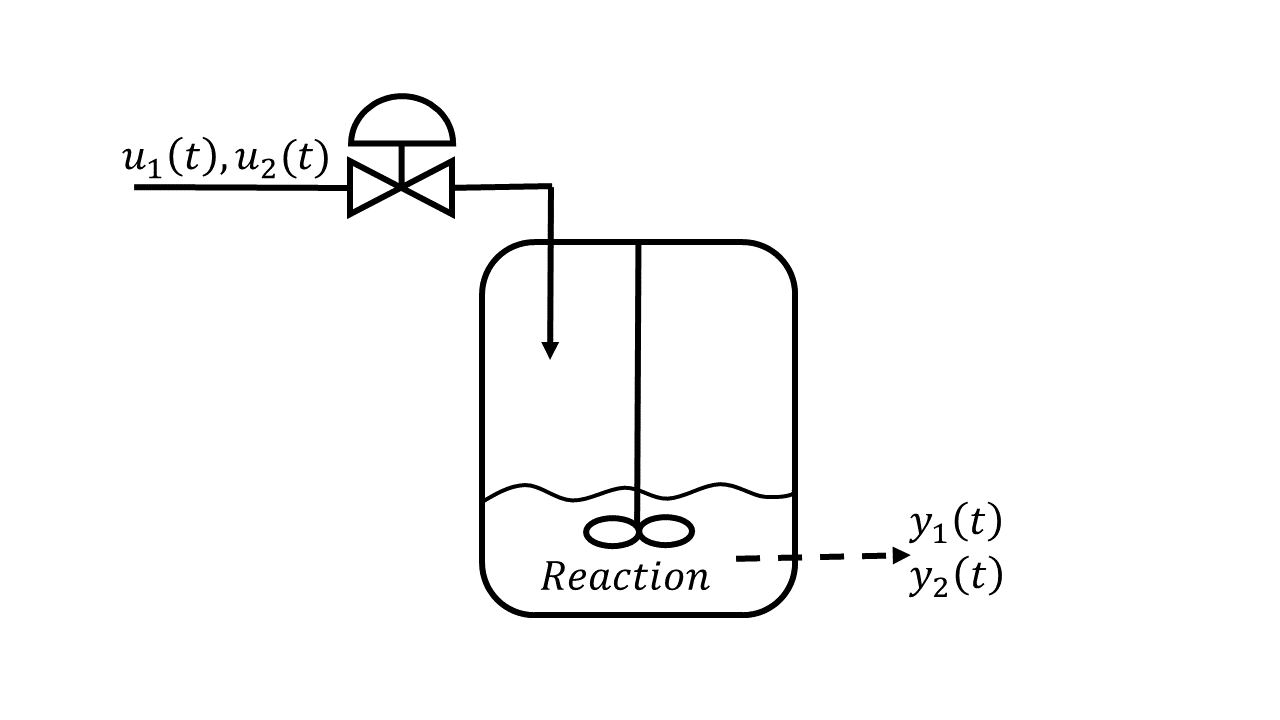}
\caption{Scheme of the yeast fermentation.}
\label{fig:YeastReactor}
\end{figure}

Yeast and a substrate are put into a reactor and the yeast ferments. Thus the substrate concentration $y_2$ decreases over time, while the biomass concentration $y_1$ increases. During this process we add additional substrate into the reactor via an input feed. This feed is governed by two (time-dependent) controls $u_1$ and $u_2$. Here $u_1$ denotes the dilution factor while $u_2$ denotes the substrate concentration of the input feed. A depiction of the setup is given in Figure \ref{fig:YeastReactor}.

Mathematically, the reaction is governed by the system of equations 

\begin{equation}\label{eq:yeast}
\begin{aligned}
\frac{dy_1(t)}{dt} & = (r(t) - u_1(t) - \theta_4)\cdot y_1(t) \\
\frac{dy_2(t)}{dt} & = - \frac{r(t) \cdot u_1(t)}{\theta_3} + u_1(t) \cdot \left( u_2(t) - y_2(t) \right) \\
r(t) & = \frac{\theta_1 \cdot y_2(t)}{\theta_2 + y_2(t)}.
\end{aligned}
\end{equation}
The time $t$ is given in hours $h$. We solve the differential equations inside the CHEMADIS Software by a $4$-th order Runge-Kutta Method with end time $t_{end} = 20\ h$.

\begin{figure}[H]
\centering
\includegraphics[width=0.45 \linewidth]{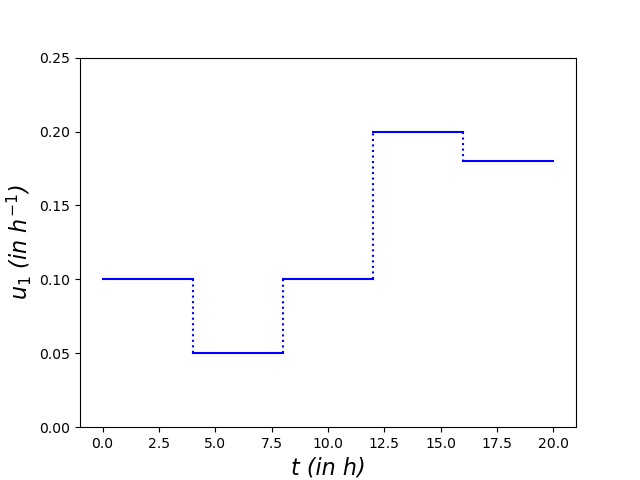}
\includegraphics[width=0.45 \linewidth]{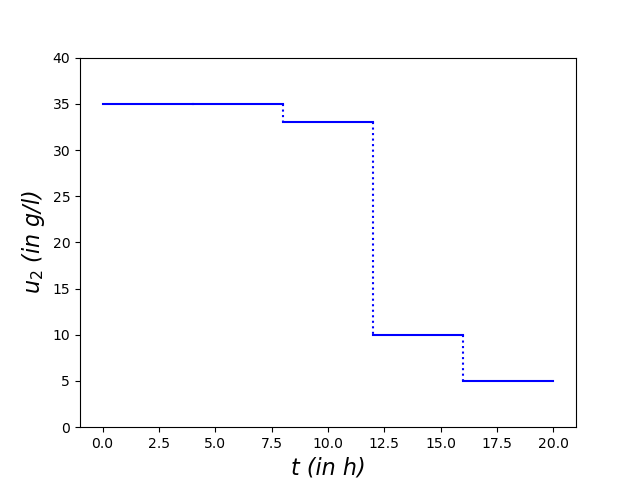}
\caption{Example of the parametrized input functions (\textbf{a}) $u_1(t)$ and (\textbf{b}) $u_2(t)$ for the  yeast fermentation model.}
\label{fig:ExampleU}
\end{figure}

In order to obtain a DoE setup we parametrize the dynamical system and replace the time-dependent functions by time-independent parameters. As inputs we consider the functions $u_1(t), u_2(t)$ and the initial biomass concentration $y_1^0 = y_1(0)$. The functions $u_i$ with $i=1,2$ are modelled as step functions with values $u_i(t) = u_{ij}$ for $t \in \left[4j,(4+1)j \right[\ h$, where $j = 0,\ldots,4$. This results in a $11$-dimensional design space with design points 

\begin{equation*}
    x = \left( y_1^0, u_{10}, \ldots, u_{14}, u_{20}, \ldots, u_{24}  \right)^T.
\end{equation*}
We bound the initial biomass concentration $y_1^0$ by $1$ \nicefrac{g}{l} and $10$ \nicefrac{g}{l}, the dilution factor $u_{1j}$ by the range $0.05$ to $0.2\ h^{-1}$ and the substrate concentration of the feed $u_{2j}$ by $5$ \nicefrac{g}{l} and $35$ \nicefrac{g}{l}. In Figure \ref{fig:ExampleU} an example of the parametrized functions $u1$ and $u2$ is plotted.

As outputs we take measurements of the biomass concentration $y_1$ in \nicefrac{g}{l} and the substrate concentration $y_2$ in \nicefrac{g}{l}. These measurements are taken at the $10$ time points $t_j^y = 2 j + 2 \ h$ for $j = 0, \ldots, 9$, each. Thus we obtain a $20$-dimensional output vector

\begin{equation*}
    y = \left( y_1(t_0^y), \ldots, y_1(t_9^y), y_2(t_0^y), \ldots, y_2(t_9^y) \right)^T.
\end{equation*}

The model parameters are $\theta_1\ \text{-}\ \theta_4$, for which we insert our current best estimate $\theta_i = 0.5$ for $i=1,\ldots,4$. This leaves one degree of freedom in the model, the initial substrate concentration which we set as $y_2(0) = 0.1$ \nicefrac{g}{l}.

For the VDM and the YBT Algorithm we introduce the grid 

\begin{equation*}
    X_{grid} = \{ 1 \nicefrac{g}{l},10 \nicefrac{g}{l} \} \times \{0.05 h^{-1}, 0.2 h^{-1} \}^5 \times \{5 \nicefrac{g}{l},20 \nicefrac{g}{l},35 \nicefrac{g}{l} \}^5
\end{equation*}
consisting of $15552$ design points. As the design space $X$ has $11$ dimensions, this grid is very coarse, despite the large amount of points. The designs $\Design_{VDM}$ and $\Design_{YBT}$ computed with these Algorithms are given in Table \ref{tab:YeastYBT}. Candidate points with a weight smaller than $0.001$ are not listed.

\begin{table}[H]
\caption{Optimal design $\Design_{VDM}$ computed with the VDM for the yeast fermentation.}
\label{tab:YeastVDM}
\centering
\tablesize{\footnotesize}
\begin{tabular}{ccccccccccccc}
\toprule
\textbf{}	& \textbf{$y_1^0$}	& \textbf{$u_{10}$} & \textbf{$u_{11}$} & \textbf{$u_{12}$} & \textbf{$u_{13}$} & \textbf{$u_{14}$} & \textbf{$u_{20}$} & \textbf{$u_{21}$} & \textbf{$u_{22}$} & \textbf{$u_{23}$} & \textbf{$u_{24}$} & \textbf{Weight}\\
\midrule
 1 & 10 \nicefrac{g}{l} & 0.05 $h^{-1}$ & 0.05 $h^{-1}$ & 0.05 $h^{-1}$ & 0.05 $h^{-1}$ & 0.05 $h^{-1}$ & 5 \nicefrac{g}{l} & 35 \nicefrac{g}{l} & 35 \nicefrac{g}{l} & 35 \nicefrac{g}{l} & 5 \nicefrac{g}{l} & 0.2445 \\
 2 & 10 \nicefrac{g}{l} & 0.2 $h^{-1}$ & 0.05 $h^{-1}$ & 0.05 $h^{-1}$ & 0.05 $h^{-1}$ & 0.05 $h^{-1}$ & 20 \nicefrac{g}{l} & 20 \nicefrac{g}{l} & 20 \nicefrac{g}{l} & 20 \nicefrac{g}{l} & 5 \nicefrac{g}{l} & 0.1111 \\
 3 & 10 \nicefrac{g}{l} & 0.2 $h^{-1}$ & 0.05 $h^{-1}$ & 0.05 $h^{-1}$ & 0.05 $h^{-1}$ & 0.05 $h^{-1}$ & 35 \nicefrac{g}{l} & 35 \nicefrac{g}{l} & 35 \nicefrac{g}{l} & 35 \nicefrac{g}{l} & 5 \nicefrac{g}{l} & 0.4517 \\
 4 & 10 \nicefrac{g}{l} & 0.2 $h^{-1}$ & 0.05 $h^{-1}$ & 0.05 $h^{-1}$ & 0.05 $h^{-1}$ & 0.05 $h^{-1}$ & 35 \nicefrac{g}{l} & 5 \nicefrac{g}{l} & 35 \nicefrac{g}{l} & 20 \nicefrac{g}{l} & 5 \nicefrac{g}{l} & 0.1920 \\
\bottomrule
\end{tabular}
\end{table}

\begin{table}[H]
\caption{Optimal design $\Design_{YBT}$ computed with the YBT Algorithm for the yeast fermentation.}
\label{tab:YeastYBT}
\centering
\tablesize{\footnotesize}
\begin{tabular}{ccccccccccccc}
\toprule
\textbf{}	& \textbf{$y_1^0$}	& \textbf{$u_{10}$} & \textbf{$u_{11}$} & \textbf{$u_{12}$} & \textbf{$u_{13}$} & \textbf{$u_{14}$} & \textbf{$u_{20}$} & \textbf{$u_{21}$} & \textbf{$u_{22}$} & \textbf{$u_{23}$} & \textbf{$u_{24}$} & \textbf{Weight}\\
\midrule
 1 & 10 \nicefrac{g}{l} & 0.05 $h^{-1}$ & 0.05 $h^{-1}$ & 0.05 $h^{-1}$ & 0.05 $h^{-1}$ & 0.05 $h^{-1}$ & 5 \nicefrac{g}{l} & 35 \nicefrac{g}{l} & 35 \nicefrac{g}{l} & 35 \nicefrac{g}{l} & 5 \nicefrac{g}{l} & 0.2446 \\
 2 & 10 \nicefrac{g}{l} & 0.2 $h^{-1}$ & 0.05 $h^{-1}$ & 0.05 $h^{-1}$ & 0.05 $h^{-1}$ & 0.05 $h^{-1}$ & 20 \nicefrac{g}{l} & 20 \nicefrac{g}{l} & 20 \nicefrac{g}{l} & 20 \nicefrac{g}{l} & 5 \nicefrac{g}{l} & 0.1113 \\
 3 & 10 \nicefrac{g}{l} & 0.2 $h^{-1}$ & 0.05 $h^{-1}$ & 0.05 $h^{-1}$ & 0.05 $h^{-1}$ & 0.05 $h^{-1}$ & 35 \nicefrac{g}{l} & 35 \nicefrac{g}{l} & 35 \nicefrac{g}{l} & 35 \nicefrac{g}{l} & 5 \nicefrac{g}{l} & 0.4520 \\
 4 & 10 \nicefrac{g}{l} & 0.2 $h^{-1}$ & 0.05 $h^{-1}$ & 0.05 $h^{-1}$ & 0.05 $h^{-1}$ & 0.05 $h^{-1}$ & 35 \nicefrac{g}{l} & 5 \nicefrac{g}{l} & 35 \nicefrac{g}{l} & 20 \nicefrac{g}{l} & 5 \nicefrac{g}{l} & 0.1921 \\
\bottomrule
\end{tabular}
\end{table}

The ADA-GPR is initiated with $200$ design points. The design computed is given in Table \ref{tab:YeastADAGPR}. Again we do not list points with a weight smaller than $0.001$ and perform the clustering described in Section \ref{sec:Materials}.

\begin{table}[H]
\caption{Optimal design $\Design_{ADAGPR}$ computed with the novel ADA-GPR for the yeast fermentation.}
\label{tab:YeastADAGPR}
\centering
\tablesize{\scriptsize}
\begin{tabular}{ccccccccccccc}
\toprule
\textbf{}	& \textbf{$y_1^0$}	& \textbf{$u_{10}$} & \textbf{$u_{11}$} & \textbf{$u_{12}$} & \textbf{$u_{13}$} & \textbf{$u_{14}$} & \textbf{$u_{20}$} & \textbf{$u_{21}$} & \textbf{$u_{22}$} & \textbf{$u_{23}$} & \textbf{$u_{24}$} & \textbf{Weight}\\
\midrule
1 & 10 \nicefrac{g}{l} & 0.1805 $h^{-1}$ & 0.05 $h^{-1}$ & 0.05 $h^{-1}$ & 0.05 $h^{-1}$ & 0.05 $h^{-1}$ & 35 \nicefrac{g}{l} & 35 \nicefrac{g}{l} & 35 \nicefrac{g}{l} & 35 \nicefrac{g}{l} & 5 \nicefrac{g}{l} & 0.3594\\
2 & 10 \nicefrac{g}{l} & 0.05 $h^{-1}$ & 0.1031 $h^{-1}$ & 0.05 $h^{-1}$ & 0.05 $h^{-1}$ & 0.05 $h^{-1}$ & 5 \nicefrac{g}{l} & 35 \nicefrac{g}{l} & 35 \nicefrac{g}{l} & 35 \nicefrac{g}{l} & 5 \nicefrac{g}{l} & 0.2543 \\
3 & 7.7720 \nicefrac{g}{l} & 0.2 $h^{-1}$ & 0.1227 $h^{-1}$ & 0.05 $h^{-1}$ & 0.05 $h^{-1}$ & 0.05 $h^{-1}$ & 35 \nicefrac{g}{l} & 35 \nicefrac{g}{l} & 35 \nicefrac{g}{l} & 23.9587 \nicefrac{g}{l} & 5 \nicefrac{g}{l} & 0.3860 \\
\bottomrule
\end{tabular}
\end{table}

As for the flash we also give a detailed breakdown of the objective value, number of iterations and the runtime in Tables \ref{tab:YeastObjective} and \ref{tab:YeastRuntime}.

\begin{table}[H]
\caption{Objective values, runtime and number of iterations of all three algorithms for the yeast fermentation.}
\label{tab:YeastObjective}
\centering
\tablesize{\small}
\begin{tabular}{cccc}
\toprule
\textbf{}	&  \textbf{VDM} & \textbf{YBT Algorithm} & \textbf{ADA-GPR}\\
\midrule
Objective value (maximization) & $8.0332$  & $8.0339$ & $8.7029$\\
Iterations & $10000$ & $5$ & $261$  \\ 
Evaluations of the Jacobian & $15552$ & $15552$ &$409$\\ 
Runtime & $115265.53$ s & $111108.58$ s & $4939.24$ s \\ 
\bottomrule
\end{tabular}
\end{table}

\begin{table}[H]
\caption{Detailed breakdown of the runtime of all algorithms for the yeast fermentation.}
\label{tab:YeastRuntime}
\centering
\begin{tabular}{cccc}
\toprule
\textbf{Runtime} & \textbf{VDM}	& \textbf{YBT Algorithm} & \textbf{ADA-GPR}\\
\midrule
Total & $115265.53$ s & $111108.58$ s & $4939.24$ s \\
Jacobian evaluation & $111106.29$ s &  $111106.29$ s &$2851.94$ s\\
Optimization: weights & - & $0.10$ s & $7.98$ s\\ 
Optimization: acquisition function & $4151.50$ s & $2.11$ s & $392.89$ s\\ 
Optimization: hyper-parameters & - & - & $1660.15$ s \\ 
\bottomrule
\end{tabular}
\end{table}

\section{Discussion}\label{sec:Discussion}

In this section we discuss the numerical results presented in Section \ref{sec:Examples}. Both examples presented differ greatly in complexity and input dimension and are discussed separately.

For the flash we observe, that the ADA-GPR can compute near optimal designs in significantly less time than the state-of-the-art YBT algorithm. In particular we need less evaluations of the Jacobian $D_\theta f$ and can drastically reduce the time required for these evaluations. This is due to the fact that the ADA-GPR operates on the continuous design space and uses an adaptive sampling instead of a pre-computed fine grid. The time reduction is also noticeable in the total runtime. Despite requiring additional steps like the hyper-parameter optimization of the GPR as well as requiring more iterations before the algorithm terminates, the ADA-GPR is faster than the YBT algorithm. The runtime needed is reduced by a factor greater than $10$.


We also see that the adaptive sampling can correctly identify points of interest for both the methanol-water as for the methanol-acetone input feed. For the former, optimal design points consist of a molar concentration $x_M < 0.3$ \nicefrac{mol}{mol}, whereas for the later the optimal concentrations also take values up to $x_M \approx 0.77$ \nicefrac{mol}{mol}. Despite using the same initial points and the same underlying system of equations the ADA-GPR identifies near optimal design points in both cases.

We observe that the VDM and the YBT algorithm compute designs with a larger objective value. This occurs as we use an approximation in the ADA-GPR instead of the exact function values $\phi(\Design,x)$ and therefore expect to have small errors in our computations. In particular for low dimensional design spaces - where the sampling of a fine grid is possible - we expect the grid based methods to result in better objective values. However, from a practical point of view, this difference is expected to be negligible. 

For the yeast fermentation we make a similar observation. The ADA-GPR can significantly reduce the number of evaluations of the Jacobian $D_\theta f$ as well as the runtime. In contrast to the flash, the ADA-GPR also computes a design with a larger - and thereby considerably better - objective value than the VDM and the YBT algorithm. 

As the design space is eleven-dimensional the grid $X_{grid}$ consisting of $15552$ design points is still very coarse. The computation of the Jacobians $D_\theta f$ for these points however takes long - more than $30$ hours. As we have a coarse grid, we do not expect the designs to be optimal on the continuous design space. In comparison, the ADA-GPR operates on the continuous design space and selects the next candidate points based on the existing information. We see, that the adaptive sampling in the continuous space leads to a better candidate point set than the arbitrary coarse grid. Using a finer grid for the VDM and the YBT Algorithm is however not possible, as the computations simply take too long. 

We conclude that the ADA-GPR outperforms a state-of-the-art algorithm for models with high-dimensional design spaces where sampling on a fine grid is computationally not tractable. Additionally, the ADA-GPR computes near optimal designs for models with low-dimensional design spaces in less time. The algorithm is particular useful for dynamical models, where a parametrization of the dynamic components can lead to many new design variables. We hereby make use of a adaptive point selection based on information based on the current candidate points instead of selecting an arbitrary fixed grid.

Still, the new algorithm also leaves room for improvement. In Tables \ref{tab:MeOHWRuntime}, \ref{tab:MeOHAceRuntime} and \ref{tab:YeastRuntime} we see that we can reduce the runtime contributed to the evaluations of the Jacobian. The runtimes for the optimization of the acquisition function as well as the optimization of the hyper-parameters however increase compared to the YBT Algorithm. In future work we want to consider and improve these steps of the algorithm.

Both VDM and YBT Algorithm have a stopping criterion which gives an error bound on the objective value, see \cite{FederovBook}. For the ADA-GPR we have no such criterion and instead use a heuristic as termination criterion - see section \ref{sec:Materials} for a detailed description. As we are using an approximation in the computations we cannot obtain an exact error bound similar to VDM and YBT Algorithm. In future work we want to investigate the quality of the computed designs to have an indication on the optimal value.

Last, we want to extend the ADA-GPR to other design of experiment settings. These include incorporating existing experiments and considering robust designs instead of locally optimal designs.

\section{Materials and Methods}\label{sec:Materials}

In this section we present details on our implementations of the VDM, the YBT Algorithm and the novel ADA-GPR which were introduced in Sections \ref{sec:Theory} and \ref{sec:ADAGPR}. We have implemented all methods using \emph{python} \cite{Python}. The models $f$ were evaluated using CHEMASIM and CHEMADIS, the BASF in-house programs (Version $6.6$ \cite{ChemasimAspiron}).

We begin by describing the grid-based VDM and YBT Algorithm. For both methods we take a grid $X_{grid}$ as input consisting of at least $d_\Theta+1$ points, where $d_\Theta$ denotes the number of unknown parameters $\theta$. We select a random initial set of candidate points $X_{n_0}$ consisting of $n_0 = d_\Theta+1$ grid points $x_i \in X_{grid}$. This amount of points is suggested in \cite{YBTAlgorithmus}, a larger amount is possible as well and can increase the numeric stability.

In the VDM we assign each candidate points $x_1, \ldots, x_{n_0}$ the weight $w_i = \frac{1}{n_0}$ in order to obtain the initial design $\Design_{n_0}$. In the YBT Algorithm we instead solve the optimal weights problem 

\begin{equation*}
    \begin{aligned}
      \min_{w_i}\ &  \Criteria\left( \sum_{i=1}^{n_0} w_i \cdot \mu(x_i) \right) \\
     s.t.\quad & \sum_{i=1}^{n_0} w_i = 1, \ 0 \leq w_i.
    \end{aligned}
\end{equation*}
We solve this problem by reformulating it as a SDP and use the \emph{mosek} Software \cite{mosek} to solve the SDP. We assign the optimal weights $w^*_i$ to the candidate points $x_i \in X_{n_0}$ to obtain the initial design $\Design_{n_0}$.

Should the Fisher Information Matrix $\Fim(\Design_{n_0})$ of the initial design $\Design_{n_0}$ be singular, we discard the design and the candidate points $X_{n_0}$ and select a new set of random candidate points.

Next we describe how we have implemented the iterative step of each algorithm. In order to obtain 

\begin{equation}
    x_{n+1} = \arg \min_x \phi(\Design_n,x)
\end{equation}
we evaluate the function $\phi(\Design_n,x)$ for every grid point $x \in X_{grid}$. We then add $x_{n+1}$ to the set of candidate points $X_n$ and adjust the weights. For the VDM we assign the new candidate point $x_{n+1}$ the weight $w_{n+1} = \frac{1}{n+1}$. The weights $w_i$ of all the previous candidate points are adjusted by multiplying with the factor $1-\frac{1}{n+1}$, resulting in the update 

\begin{equation*}
    w_i \to w_i \cdot\left( 1-\frac{1}{n+1} \right).
\end{equation*}
This factor is chosen according to \cite[Chapter 3.1.1]{FederovBook}. In the YBT Algorithm we instead adjust the SDP to also account for the new candidate point $x_{n+1}$ and re-solve the weight optimization problem. With the updated weights we obtain the design $\Design_{n+1}$ and can iterate. 

Last we discuss our stopping criterion. We set a value $\varepsilon = 10^{-3}$ and stop the algorithm as soon as $\min \phi(\Design_n,x) > -\varepsilon$. The computed design $\Design_n$ then fulfills $\Criteria\left( \Fim(\Design_n) \right) - \min_{\Design \in \DESIGN{X}} \Criteria\left( \Fim(\Design) \right) < \varepsilon$. Setting a smaller value of $\varepsilon$ increases the precision of the design, but also increases the number of iterations needed. Additionally we terminate the algorithm if we reach $10000$ iterations.

Now we discuss our implementation of the novel ADA-GPR. As we want to use a Gaussian process regression, it is helpful to scale the inputs. We thus map the design space to the unit cube $[0,1]^{d_X}$.

In the ADA-GPR we select a number $n_0 > d_\Theta$ of initial candidate points. For the examples from Section \ref{sec:Examples} we have selected $50$ and $200$ initial points respectively. The initial points are set as the first $n_0$ points of the $d_X$-dimensional Sobol-sequence \cite{Sobol1967,Sobol1976}. This is a pseudo-random sequence which uniformly fills the unit cube $[0,1]^{d_X}$. In our experience, one has to set $n_0$ significantly larger than $d_\Theta$, on the one hand to ensure the initial Fisher Information Matrix is not singular, on the other hand to obtain a good initial approximation. We obtain the weights for the candidate points $X_{n_0}$ analogously to the YBT Algorithm by solving the SDP formulation of the optimal weights problem with \emph{mosek}.

Next we compute a Gaussian process regression for the directional derivative $\phi(\Design_n,x)$ based on the evaluations $\left( X_n,\phi(\Design_n,X_n)\right)$. For this GPR we use the machine learning library \emph{scikit-learn} \cite{scikit-learn} with the squared exponential kernel \emph{RBF}. The kernel is dependent on $3$ hyper-parameters, a pre-factor $\sigma_f^2$, the lengthscale $l$ and a regularity factor $\alpha$. The parameters $\sigma_f^2$ and $l$ are chosen via the \emph{scikit-learn} built-in loss function. They are chosen every time we fit the Gaussian process to the data, i.e. in every iteration. For the factor $\alpha$ we use cross-validation combined with a grid search, where we consider the $21$ values $\alpha = 10^{-10} , 10^{-9.5}, 10^{-9}, \ldots, 10^{0}$. As the cross validation of the hyper-parameters can be time-expensive, we do not perform this step in every iteration. Instead we adjust $\alpha$ in the first $n_{init} = 10$ iterations and then only every $10$th iteration afterwards.

Now we discuss the optimization of the acquisition function

\begin{equation*}
    \E{\phi(\Design_n,x)\left| X_n, \phi(\Design_n,X_n) \right.} - \Var{\phi(\Design_n,x)\left| X_n, \phi(\Design_n,X_n) \right.}.
\end{equation*}
In order to obtain a global optimum we perform a multistart, where we perform several optimization runs from different initial values. In our implementation we perform $n_{nr\_opt} = 10$ optimization runs. The initial points are selected via the $d_X$-dimensional Sobol-sequence in the design space $[0,1]^{d_X}$. We recall, that the Sobol-sequence was also used to select the initial candidate points. In order to avoid re-using the same points, we store the index of the last Sobol-point we use. When selecting the next batch of Sobol-points, we take the points succeeding the stored index. Then we increment the index.
For the optimization we use the \emph{L-BFGS-B} method from the \emph{scipy.optimize} library \cite{2020SciPy,L_BFGS}.

Last we present the stopping heuristic we use for the novel ADA-GPR. Throughout the iterations we track the development of the objective value and use the progress made as stopping criterion for the algorithm. For the initial $50$ iterations, we do not stop the algorithm. After the initial iterations we consider the progress made over the last $40\%$ of the total iterations. However we set a maximum of $50$ iterations which we consider for the progress. For the current iteration $n_{cur}$ we thus compute $n_{stop} = \max(0.6 \cdot n_{cur}, n_{cur}-50)$ and consider the progress

\begin{equation*}
    \Delta \Criteria_{stop}= \Criteria(\Fim(\Design_{n_{cur}})) - \Criteria(\Fim(\Design_{n_{stop}})).
\end{equation*}
If $\Delta \Criteria_{stop} < 0.001$, we stop the computation. Else we continue with the next iteration.

For the tables and figures from Section \ref{sec:Examples} we have clustered the results from the ADA-GPR. Here we have proceeded as described in the following. We iterate through the support points of the computed design $\Design_{ADAGPR}$. Here we denote these support points by $s_i$. For each point $s_i$ we check if a second distinct point $s_j$ exists, such that $\| s_i - s_j \| < 0.01$. If we find such a pair, we add these points to one joint cluster $C_i$. If we find a third point $s_k$ with either $\| s_k - s_j \| < 0.01$ or $\| s_k - s_i \| < 0.01$, the point $s_k$ is added to the cluster $C_i$ as well. 

If for a point $s_i$ no point $s_j$ exists such that $\| s_i - s_j \| < 0.01$, the point $s_i$ initiates its own cluster $C_i$.

After all points are divided into clusters, we represent each cluster $C_i$ by a single point $c_i$ with weight $w^c_i$. The point $c_i$ is selected as average over all points $s_j$ in the cluster $C_i$ via the formula 

\begin{equation*}
    c_i = \frac{1}{|C_i|} \sum_{s_j \in C_i} s_j.
\end{equation*}
The weight $w^c_i$ is set as sum of the weights assigned to the points in the cluster $C_i$ and is computed via 

\begin{equation*}
    w^c_i =  \sum_{s_j \in C_i} w_j.
\end{equation*}

\vspace{6pt} 



\authorcontributions{Conceptualization, P.S, J.S. and M.B.; methodology, P.S, J.S. and M.B.; software, P.S.; validation, P.S. and M.B.; formal analysis, P.S.; investigation, P.S., J.S and M.B.; resources, J.S. and M.B.; data curation, P.S.; writing--original draft preparation, P.S.; writing--review and editing, J.S. and M.B.; visualization, P.S.; supervision, J.S. and M.B.; project administration, M.B. All authors have read and agreed to the published version of the manuscript.}

\funding{This research received no external funding.}

\acknowledgments{The authors thank Prof. Dr. Karl-Heinz Küfer, Dr. Tobias Seidel, Dr. Charlie Vanaret and Dr. Norbert Asprion for their support in the research and the helpful discussions. They also thank Dr. Norbert Asprion and the BASF SE for access to the CHEMASIM Software, the CHEMADIS Software and the chemical engineering examples.}

\conflictsofinterest{The authors declare no conflict of interest.} 




\appendixtitles{yes} 
\appendix
\section{MESH Equations}\label{app:MESH}
The flash presented in Section \ref{sec:Examples} is governed by the so-called MESH equation. These equations are given as (see \cite{biegler1997})

\begin{itemize}[leftmargin=*,labelsep=5.8mm]
\item Mass balances

\begin{align*}
F \cdot x_m  & = V \cdot y^{vap}_m + L \cdot y^{liq}_m \\
F \cdot x_w  & = V \cdot y^{vap}_w + L \cdot y^{liq}_w
\end{align*}

\item Equilibrium

\begin{align*}
P \cdot y^{vap}_m & = P_m^0(T) \cdot y^{liq}_m \cdot \gamma_m(y^{liq}_m,y^{liq}_w, T) \\
P \cdot y^{vap}_w & = P_w^0(T) \cdot y^{liq}_w\cdot \gamma_w(y^{liq}_m,y^{liq}_w, T)
\end{align*}

\item Summation

\begin{equation*}
x_m + x_w = y^{vap}_m + y^{vap}_w = y^{liq}_m + y^{liq}_w = 1
\end{equation*}

\item Heat balance

\begin{equation*}
\dot{Q} + F \cdot H_L(x_m, x_w, T_F) = V \cdot H_V(y^{vap}_m, y^{vap}_w, T) + L \cdot H_L(y^{liq}_m, y^{liq}_w, T).
\end{equation*}

\end{itemize}

The functions $P_m^0, P_w^0$ denote the vapor pressure of the pure elements and are given as 

\begin{equation*}
    P^0_s(T) = \exp \left( A_s + \frac{B_s}{T} + C_s \ln(T) +D_s T^{E_s} \right)
\end{equation*}

for the substance $s$. The parameters of the vapor pressure $A_s$ to $E_s$ also depend on the substance. For methanol, water and acetone they are listed in Table \ref{tab:SubPara}.

\begin{table}[H]
\caption{Substance parameters for the flash.}
\label{tab:SubPara}
\centering
\begin{tabular}{cccc}
\toprule
\textbf{} & \textbf{methanol}	& \textbf{water} & \textbf{acetone}\\
\midrule
$A_s$ & 100.986 & 64.36627 & 78.89993\\
$B_s$ & -7210.917 & -6955.958 & -5980.876 \\
$C_s$ & -12.44128 & -5.802231 & -8.636991\\
$D_s$ & $1.307676 \cdot 10^{-2}$ & $3.114927\cdot 10^{-9}$ & $7.92829 \cdot 10^{-6}$ \\
$E_s$ & 1 & 3 & 2\\
\bottomrule
\end{tabular}
\end{table}

The functions $H_{liq}, H_{vap}$ denote the enthalpies of the molar liquid and vapor streams. The activity coefficients $\gamma_s$ are given by

\begin{equation*}
\begin{aligned}
    & \gamma_m(y^{liq}_m,y^{liq}_w, T) = \\
    & \exp\left((y^{liq}_w)^2 \cdot \tau_{21} \cdot\frac{\exp(-\alpha \cdot \tau_{21})^2}{(x_m + x_w \cdot \exp(-\alpha \cdot \tau_{21}) )^2} + \tau_{12} \cdot \frac{\exp(-\alpha \cdot \tau_{12})}{(x_w + x_m \cdot \exp(-\alpha \cdot \tau_{12}) )^2}\right) \\
    \end{aligned}
\end{equation*}
and 

\begin{equation*}
\begin{aligned}
    & \gamma_w(y^{liq}_m,y^{liq}_w, T) = \\
    & \exp\left((y^{liq}_m)^2 \cdot \tau_{12} \cdot\frac{\exp(-\alpha \cdot \tau_{12})^2}{(x_w + x_m \cdot \exp(-\alpha \cdot \tau_{12}) )^2} + \tau_{21} \cdot \frac{\exp(-\alpha \cdot \tau_{21})}{(x_m + x_w \cdot \exp(-\alpha \cdot \tau_{21}) )^2}\right),\\
    \end{aligned}
\end{equation*}
with 

\begin{equation*}
 \tau_{12} = a_{12}+\frac{b_{12}}{T} \quad \text{and} \quad \tau_{21} = a_{21}+\frac{b_{21}}{T}
\end{equation*} 

and where we set the parameter $\alpha$ as $\alpha = 0.3$. The NRTL parameters $a_{12}, a_{21}, b_{12}$ and $b_{21}$ are set as $(a_{12}, a_{21}, b_{12}, b_{21}) = ( -3.8, 6.6, 1337.558, -1900)$ for the methanol-water mixture and as $(a_{12}, a_{21}, b_{12}, b_{21}) = ( 4.1052, -4.4461, -1264.515, 1582.698)$ for the methanol-acetone mixture.

\reftitle{References}


\externalbibliography{yes}
\bibliography{Bibliography}






\end{document}